\documentclass[10pt,journal,letterpaper]{IEEEtran}

\usepackage{hyperref}
\usepackage{amssymb}
\usepackage{graphicx}
\usepackage{color}
\usepackage{amsmath}
\newtheorem{definition}{Definition}[section]
\newtheorem{lemma}{Lemma}[section]
\newtheorem{proposition}{Proposition}[section]

\newtheorem{example}{Example}[section]

\newtheorem{remark}{Remark}[section]

\newtheorem{conjecture}{Conjecture}[section]

\usepackage{graphicx}
\usepackage{verbatim}
\usepackage{url}
\usepackage{amsmath}
\usepackage{amsfonts}
\usepackage{amssymb}
\usepackage{enumerate}
\usepackage{hyperref}

\newcommand{\argmin}{\operatornamewithlimits{argmin}}
\newcommand{\argmax}{\operatornamewithlimits{argmax}}
\newcommand{\PoA}{{\rm PoA}}
\newcommand{\remove}[1]{}

\begin{document}
\title{Combined Base Station Association and Power Control in Multi-channel Cellular Networks}

\author{Chandramani~Singh,~\IEEEmembership{Student Member,~IEEE,}
Anurag~Kumar,~\IEEEmembership{Fellow,~IEEE,} \\
and~Rajesh~Sundaresan,~\IEEEmembership{Senior Member,~IEEE}
\thanks{This is an extended version of a paper that appeared in Gamenets 2009.}
\thanks{This work was supported by an INRIA Associates program DAWN, and also by the Indo-French Centre for the Promotion of Advanced Research (IFCPAR), Project No.~4000-IT-A.}
\thanks{The authors are with the Department of Electrical Communication Engineering
Indian Institute of Science Bangalore, India~(email: \{chandra,~anurag,~rajeshs\}@ece.iisc.ernet.in).}
}

\maketitle

\begin{abstract}
A combined base station association and power control problem is
studied for the uplink of multichannel multicell
cellular networks, in which each channel is used by exactly one cell
(i.e., base station). A distributed association and power update
algorithm is proposed and shown to converge to a Nash equilibrium of a
noncooperative game. We consider network models with discrete mobiles
(yielding an atomic congestion game), as well as a continuum of mobiles (yielding a
population game). We find that the equilibria need not be Pareto efficient, nor
need they be system optimal. To address the lack of system optimality,
we propose  pricing mechanisms. It is shown that these mechanisms can be
implemented in a distributed fashion.
\end{abstract}

\section{Introduction}
Wireless communication systems have experienced tremendous growth
over the last decade, and this growth continues unabated worldwide.
Efficient management of resources is essential for the success of
wireless cellular systems. In a mobile cellular system, mobiles
adapt to time varying radio channels by adjusting base station~(BS)
associations and by controlling transmitter powers. Doing so, they
not only maintain their quality of service~(QoS) but also enhance
their transmitters' battery lives. In addition, such controls reduce
the network interference, thus maximizing spatial spectrum reuse.
Distributed control is of special interest, since the alternative of
centrally orchestrated control involves added infrastructure, the
need for distribution of measurements, and hence system complexity.

Distributed control algorithms for single channel multicell networks have
been extensively studied~(Foschini \& Miljanic~\cite{ctrltheory-cellular.foschini-miljanic93distributed-power-control},
Yates~\cite{ctrltheory-cellular.yates95uplink-power-control}, Hanly~\cite{ctrltheory-cellular.hanly95cell-site-selection-power-control}).
The monograph by Chiang
et~al.~\cite{ctrltheory-cellular.chiang-etal08power-control-cellular-networks}
and references therein provide an excellent survey of the area.
Noncooperative games have been a natural tool for analysis and
design of distributed power control algorithms. Scutari
et~al.~\cite{gametheory-cellular.scutari-etal06potential-games-vector-power-control}
and
Heikkinen~\cite{gametheory-cellular.heikkinen06potential-game-power-control-scheduling}
model distributed power control problems as potential games, while
Altman \&
Altman~\cite{gametheory-wireless.altman-altman03s-modular-games}
show that many of the cellular power control algorithms can be
modeled as submodular games. In contrast, uplink resource allocation
for {\it multichannel} multicell networks poses several challenges
as observed in
Yates~\cite{ctrltheory-cellular.yates95uplink-power-control} and
Jiang
et~al.~\cite{gametheory-cellular.jiang-etal08base-station-association-game}.

We address the resource allocation problem in the uplink of a
multichannel multicell network with a single traffic class. Such a
problem arises when a CDMA operator chooses to lease and utilize
multiple frequency bands in order to reduce in-network interference,
or multiple operators who lease different bands decide to cooperate.
The newer mobile devices are typically radio agile, and thus have
the option to choose from one of these distinct bands.
We address a simplified version of this multichannel multicell
problem where all BSs operate on different frequency bands.

A preview of our results is as follows. We propose a distributed
algorithm for the combined base station association and power
control problem, and subsequently  model the problem as a
player-specific congestion game. The equilibrium states of such
algorithms, which are Nash equilibria of the corresponding games,
may be far from system optimum. We resort to pricing mechanisms to
induce mobiles to behave in a way that optimizes system cost. We
also show that such a mechanism can be employed in a distributed
fashion.
Towards this end, we model the network as having a continuum of~(nonatomic) mobiles, each offering
infinitesimal load, which leads to a population game formulation.
We then provide a marginal pricing mechanism that
motivates a pricing strategy for the discrete mobiles case.
Note that, unlike the case of transportation networks, mobiles are not
really priced in cellular networks. The pricing is simply a part of the decision making
routine built into each mobile in order bring about a {\em distributed} control mechanism
that drives the system toward optimality.

The paper is organized as follows. In Section~\ref{game-pre} we briefly
discuss concepts of {\it finite noncooperative games} and {\it population games}.
We study a network model with discrete mobiles in Section~\ref{disc-users}. We propose a combined association
and power control algorithm, model it as a noncooperative game,
and analyze its performance. We extend this analysis to a network with a continuum of mobiles in
Section~\ref{cont-users}. To address the inefficiency of the proposed algorithms, we
design toll mechanisms in Section~\ref{pricing}. Finally, we conclude the paper with some remarks in
Section~\ref{future-work}.

Optimal power allocation and BS association in uplinks of
multichannel multicell cellular networks have not been explored
before. Ours is an attempt at a detailed coverage on what is
possible in general, with more specific results in some special
cases.

\section{Game Preliminaries}

\label{game-pre}
\subsection{Finite Noncooperative Games}
\label{games-finite}
A noncooperative strategic form game $(\mathcal{M}, (\mathcal{A}_i, i \in \mathcal{M}), (c_i, i \in \mathcal{M}))$
consists of a set of players $\mathcal{M} = \{1,\dots, M\}$.
Each player $i$ is accompanied by an action set $\mathcal{A}_i$ and a cost function
$c_i:\times_{i = 1}^M \mathcal{A}_i \rightarrow \mathbb{R}$. In this work, we
assume all action sets to be finite. An action profile $\mathbf{a} = (a_i, i = 1,\dots,M)$
prescribes an action $a_i$ for every player $i \in \mathcal{M}$.
For $\mathbf{a} = (a_i, i = 1,\dots,M)$, denote $\mathbf{a}_{-i} := (a_1,\dots,a_{i-1},a_{i+1},\dots,a_M)$
and $(b_i,\mathbf{a}_{-i}) := (a_1,\dots,a_{i-1},b_i,a_{i+1},\dots,a_M)$.
\begin{definition}
Nash Equilibrium~(NE): For an action profile $\mathbf{a}$, a mobile
$i$'s \emph{best response}, $\mathcal{B}_i(\mathbf{a}) \subseteq
\mathcal{A}_i$, is defined as $\mathcal{B}_i(\mathbf{a}) :=
\argmin_{b_i \in \mathcal{A}_i} c_i(b_i, a_{-i})$. $\mathbf{a}$ is
said to be a Nash Equilibrium for the game if $a_i \in
\mathcal{B}_i(\mathbf{a})$ for all $i \in \mathcal{M}$.
\end{definition}

\begin{definition}
Potential Game: A game $(\mathcal{M}, (\mathcal{A}_i, i \in
\mathcal{M}), (c_i, i \in \mathcal{M}))$  is said to be an ordinal
potential game if there exists a function $V : \times_{i = 1}^M
\mathcal{A}_i \rightarrow \mathbb{R}$, known as an {\it ordinal
potential function}, that satisfies
\[c_i(b_i,\mathbf{a}_{-i}) < c_i(\mathbf{a}) \Leftrightarrow  V(b_i,\mathbf{a}_{-i}) < V(\mathbf{a})\]
for all $i \in \mathcal{M}, b_i \in \mathcal{A}_i, \mathbf{a} \in \times_{i = 1}^M \mathcal{A}_i$.
\remove{
\item a best-response potential game if there exists
a function $V : \times_{i = 1}^M \mathcal{A}_i \rightarrow \mathbb{R}$, known as a best-response
potential function, that satisfies
\[\argmin_{b_i \in \mathcal{A}_i}c_i(b_i,\mathbf{a}_{-i}) = \argmin_{b_i \in \mathcal{A}_i} V(b_i,\mathbf{a}_{-i})\]
for all $i \in \mathcal{M}, b_i \in \mathcal{A}_i, \mathbf{a} \in \times_{i = 1}^M \mathcal{A}_i$
\end{enumerate}
}
\end{definition}

\remove{ Clearly all ordinal potential functions are also
best-response potential functions, and so, all ordinal potential
games are also best-response potential games. The reverse
implications are not true in general. } Clearly, all minimizers of
an ordinal  potential function $V$ are Nash equilibria of the game.
Thus all ordinal potential games $(\mathcal{M}, (\mathcal{A}_i, i
\in \mathcal{M}), (c_i, i \in \mathcal{M}))$ admit at least one Nash
equilibrium. On account of their finiteness they also have the
finite improvement path~(FIP) property, i.e., do not contain
improvement cycles~(Monderer \&
Shapley~\cite[Lemma~2.3]{gametheory.monderer-shapley96potential-games}).
Thus, in a finite ordinal potential game when players update as per
the {\it better response strategy},
 {\it round-robin} or {\it random} update processes
converge to a Nash equilibrium in a finite number of steps.
With the same strategies, an {\it asynchronous} update process
also converges~(Neel~\cite[Chapter~5]{gametheory-cogradio.neel06thesis}).
\begin{remark}
The strategic form games that have the FIP property also admit the
finite best-response path~(FBRP) property, i.e., do not contain best
response
cycles~(Milchtaich~\cite[Section~5]{gametheory.milchtaich96congestion-games-player-specific}).\footnote{A
best response cycle is a finite best response path
$\mathbf{a^1},\cdots,\mathbf{a^k}$ such that $\mathbf{a^1} =
\mathbf{a^k}$, and for some $j \in \{1,\cdots,k-1\}$, the deviating
player in iteration $j$ strictly
benefits\cite{gametheory.voorneveld00best-response-potential-games}.}
Thus, if players update as per the {\it best response} strategy,
then also the above update processes converge to a Nash equilibrium
in a finite number of steps. The reverse implication is not true in
general - FBRP need not imply FIP.
\end{remark}

\begin{definition}
Congestion Game: A game $(\mathcal{M}, (\mathcal{A}_i, i \in \mathcal{M}), (c_i, i \in \mathcal{M}))$
is said to be a player-specific weighted singleton congestion game if
\begin{enumerate}
\item there exists a set $\mathcal{N}$ such that $\mathcal{A}_i = \mathcal{N}$
for all $i \in \mathcal{M}$, and
\item there exist constants $(\beta_i, i \in \mathcal{M})$ and nonincreasing functions $f_{ij}, i \in \mathcal{M},j \in \mathcal{N}$ such that
\[c_i(\mathbf{a}) = f_{ia_i}(\sum_{\stackrel{l \in \mathcal{M}:}{a_l = a_i}}\beta_l) \mbox{ for all } \mathbf{a} \in \times_{i = 1}^M \mathcal{A}_i, i \in \mathcal{M}.\]
\end{enumerate}
\end{definition}
In the above definition, we interpret $\mathcal{N}$ as a set of facilities and $\beta_l$ as the load offered by player $i$. Then, $\sum_{\stackrel{l \in \mathcal{M}:}{a_l = a_i}}\beta_l$ denotes the total load on facility~$a_i$, under an action profile $\mathbf{a}$. The game is a {\em singleton} congestion game because each action picks exactly one facility. It is {\it weighted} because players offer different loads, and it is {\it player-specific} because the cost functions $c_i(\cdot)$ are player-specific.

Rosenthal~\cite{gametheory.rosenthal73games-pure-strategy-nash-equilibria} has defined
congestion games with unweighted players and player-independent cost functions, but more general action sets.
The above generalization is due to Milchtaich~\cite{gametheory.milchtaich96congestion-games-player-specific} who also
showed that singleton congestion games with weighted players but player independent costs admit FIP.
Gairing et al.~\cite{gametheory-comnet.gairing-etal06routing-player-specific-linear-latency} studied these games in the special
case of linear cost functions. Georgiou et al.~\cite{gametheory-comnet.georgiou-etal09network-uncertainty-routing}
introduced a routing game where the players only have incomplete information about edge latencies, and showed
that this game can be transformed into a~(complete information) player specific singleton congestion game
with linear cost functions. Mavronicolas et al.~\cite{gametheory.mavronicolas-etal07congestion-games-player-specific}
considered a subclass where each player-specific cost function is composed~(by means of an abelian
group operation) of a player-specific constant and a facility-specific nondecreasing function.


\subsection{Population Games}
\label{pop-games}
A population game~(Sandholm~\cite{gametheory.sandholm01potential-games-continuous-player-sets})
$(\mathcal{M}, (\mathcal{A}_l, l \in \mathcal{L}), (c_{lj},  l \in \mathcal{L}, j \in \mathcal{A}_l))$ consists
of $\mathcal{L} = \{1,\dots,L\}$
classes of nonatomic populations of players. $\mathcal{M} = \cup_{l \in \mathcal{L}} \mathcal{M}_l$,
and $M_l := |\mathcal{M}_l|$ denotes the total mass of the class $l$ population. By a nonatomic
population, we mean that the mass of each member of the population is infinitesimal.
Players of class $l$ are associated with an action set  $\mathcal{A}_l$. Actions of these~(class $l$) players lead to
an action distribution $\mathbf{m}^l = (m_{lj}, j \in \mathcal{A}_l)$,
where $\sum_{j \in \mathcal{A}_l}m_{lj} = M_l$. All the players within
a class are alike. Thus the action distributions completely specify the play; we can characterize the
states and dynamics of play solely in terms of action distributions.
Let $\mathbf{m} = (\mathbf{m}^l, l \in \mathcal{L})$ denote the action distribution profile
across the entire population, and let $\mathcal{M}^{\ast}$ denote the set of all such profiles.
A population $l$ is also accompanied by continuous cost density functions $c_{lj}: \mathcal{M}^{\ast} \rightarrow \mathbb{R}$.

\begin{definition}
\label{cont-player-NE}
Nash Equilibrium~(NE): An action distribution profile $\mathbf{m}$ is a pure strategy Nash equilibrium for the game $(\mathcal{M}, (\mathcal{A}_i, i \in \mathcal{M}), (c_{lj},  l \in \mathcal{L}, j \in (\mathcal{A}_l))$ if and only if for all $l \in \mathcal{L}$  and $j \in \mathcal{A}_l$, a positive mass $m_{lj} > 0$ implies  $c_{lj}(\mathbf{m}) \leq c_{lk}(\mathbf{m})$ for all $k \in \mathcal{A}_l$.
\end{definition}

\begin{remark}
\label{pop-NE}
At a Nash equilibrium $\mathbf{m}$, for a class $l$, if $j$ and $k$ are any two facilities in $\mathcal{A}_l$ such that $m_{lj} > 0, m_{lk} > 0$, then $c_{lj}(\mathbf{m}) = c_{lk}(\mathbf{m})$.
\end{remark}

\begin{definition}
\label{cont-potential}
 Potential Game: A game $(\mathcal{M},
(\mathcal{A}_l, l \in \mathcal{L}), (c_{lj},  l \in \mathcal{L}, j
\in \mathcal{A}_l))$ is said to be a potential game if there exists
a $\mathbf{C}^1$ function $V : \mathcal{M}^{\ast} \rightarrow
\mathbb{R}$, known as a {\it potential function}, that satisfies
\[\frac{\partial V(\mathbf{m})}{\partial m_{lj}} = c_{lj}(\mathbf{m})\]
for all $l \in \mathcal{L}, j \in \mathcal{A}_l,  \mathbf{m} \in \mathcal{M}^{\ast}$.
\end{definition}

It is well known that Nash equilibria are the profiles
which satisfy the Kuhn-Tucker first order conditions for a minimizer of the potential
function~(Sandholm~\cite[Proposition~3.1]{gametheory.sandholm01potential-games-continuous-player-sets}).
Any dynamics with {\it positive correlation} and
{\it noncomplacency}~(in particular the best response dynamics) approaches
a Nash equilibrium~\cite{gametheory.sandholm01potential-games-continuous-player-sets}.

We are interested in \emph{nonatomic congestion games}~(Sandholm~\cite{gametheory.sandholm01potential-games-continuous-player-sets}), in which $\mathcal{A}_l = \mathcal{N}, \forall l \in \mathcal{L}$,
for a given set $\mathcal{N}$. As before, we interpret $\mathcal{N}$ as a set of facilities.
Moreover, each class $l$ has an associated offered load density $\gamma_l > 0$.
An action distribution profile $\mathbf{m}$ leads to a congestion profile $(m_j, j \in \mathcal{N})$,
where $m_j =  \sum_{l \in \mathcal{L}}m_{lj} \gamma_l$. The cost density functions $c_{lj}$ depend
on $\mathbf{m}$ only through $m_j$, and are increasing in $m_j$.

\subsection{Pricing}
\label{pricing-literature}
Levying of tolls is a conventional way to enforce system optimality in nonatomic networks.
Beckman~\cite{gametheory.beckman-etal56economics-transportation} and
Dafermos \& Sparrow~\cite{gametheory.dafermos-sparrow69traffic-assignment} studied optimal tolls
in transportation networks with a single class of users.
Later Dafermos~\cite{gametheory.dafermos73traffic-assignment-multi-class}
and Smith~\cite{gametheory.smith79marginal-cost-taxation} extended the analysis
to multiclass networks. Roughgarden \& Tardos~\cite{gametheory-comnet.roughgarden-tardos02selfish-routing}
applied these ideas in computer networks and analyzed tolls for optimal routing.

In the atomic~(discrete) setting, Caragiannis et al.~\cite{gametheory.caragiannis-etal} proposed tolls for two-terminal parallel-edge
networks with unweighted users and linear latency functions. Fotakis \& Spirakis~\cite{gametheory.fotakis-spirakis}
extended these to generic two-terminal networks with unweighted users and arbitrary increasing
latency functions. We propose an alternative toll mechanism, and demonstrate that the proposed tolls can be computed in
a distributed fashion.

\section{Discrete Mobiles}

\label{disc-users}
\subsection{System Model}
\label{sys-model} We now describe the model adopted in this work. We
consider the uplink of a cellular network consisting of several BSs
and mobiles. {\it Each BS operates in a distinct frequency band.}
Let $\mathcal{N} = \{1,\dots, N\}$ and $\mathcal{M} = \{1,\dots,
M\}$ denote the set of BSs and the set of mobiles, respectively.

A mobile must be associated with one BS at any time, and is free to
choose the BS with which it associates. Let $h_{ij}$ denote the
power gain from mobile $i$ to BS $j$. Let the receiver noise at all
BSs have the average power $\sigma^2$.
Let $p_i$ denote the power transmitted by mobile $i$, and let $a_i$
be the BS to which it is associated. Under an association  profile
$\mathbf{a} = (a_i, i = 1, \dots, M)$, let
$\mathcal{M}_j(\mathbf{a})$ be the set of mobiles associated with
BS~$j$. Under an association profile $\mathbf{a} = (a_i, i = 1,
\dots, M)$ and a power vector $\mathbf{p} = (p_i, i = 1, \dots, M)$,
the signal to interference plus noise ratio~(SINR) of mobile $i$ at
BS $a_i$ is
\[\frac{h_{ia_i}p_i}{\sum_{l \in \mathcal{M}_{a_i}(\mathbf{a}) \setminus \{i\}} h_{la_l}p_l + \sigma^2}\]
Mobile $i$ has a target SINR requirement $\gamma_i$.


\subsection{The Proposed Algorithm}
\label{atomic-assoc}
Yates~\cite{ctrltheory-cellular.yates95uplink-power-control} and
Hanly~\cite{ctrltheory-cellular.hanly95cell-site-selection-power-control}
 proposed an algorithm for distributed association and power control
in single channel cellular networks.
Convergence results for the algorithm are based
on the concept of a {\it standard interference function}.
The technique is based on a mobile reassociating itself with a BS with which it needs to
use the least power; this fails to work in the case of a multichannel network
and analogous convergence results for this algorithm may not
hold~(see Yates~\cite[Section~VI]{ctrltheory-cellular.yates95uplink-power-control}). Even in
instances where the algorithm converges,
it may get stuck at a power allocation that is not Pareto efficient.

We propose an alternative distributed algorithm for combined BS
association and power control in multichannel multicell cellular
networks. We also show its convergence. We make use of the following
simple fact~(see, for example, Kumar
et~al.~\cite[Chapter~5]{commnet-wireless.kumar-etal08wireless-networking}).
Consider the subproblem of power control with a fixed association
$\mathbf{a}$.
Define $\beta_l = \frac{\gamma_l}{1 + \gamma_l}$, a measure of the
``load'' offered by mobile $l$.

\begin{proposition}
\label{dis-steady-pow} For a fixed association $\mathbf{a}$,\\
(i) The power control subproblem of BS~$j$ is feasible iff $\sum_{l \in \mathcal{M}_j(\mathbf{a})}\beta_l < 1$. \\
(ii) If the power control subproblem of BS~$j$  is feasible,
there exists a unique Pareto efficient power vector $\mathbf{p}(\mathbf{a})$ given by
\[p_i(\mathbf{a}) = \frac{\sigma^2}{h_{ij}} \frac{\beta_i}{1 - \sum_{l \in \mathcal{M}_j(\mathbf{a})}\beta_l}.\]
\end{proposition}

Throughout we assume that there exists at least one feasible
association and power vector. Proposition~\ref{dis-steady-pow}
motivates the following algorithm.

\noindent {\bf Multichannel Association and Power Control~(MAPC):}
Mobiles switch associations in a round-robin fashion by taking into
account the optimal power consumptions~(given by
Proposition~\ref{dis-steady-pow}(ii)) at the BSs with which these
associate. This is done as follows. As the load at a BS changes, it
immediately broadcasts the new load, and associated mobiles update
their powers to the optimal required powers as per the new loads. To
be more precise, define
\begin{equation}
\label{cost-fun}
c_i(\mathbf{a}) = \frac{\sigma^2}{h_{ia_i}} \frac{\beta}{[1 - \sum_{l \in \mathcal{M}_j(\mathbf{a})}\beta_l]^+},
\end{equation}
where $[x]^+ = \max(x,0)$. For $t = 0,1,2,\dots$, mobile $i$ where
$i = 1 + (t \mod M)$ updates its association and power at~$t+1$ if
$a_i(t) \notin \argmin_{j \in \mathcal{N}}
c_i((j,\mathbf{a}(t)_{-i}))$. In this case,
\begin{subequations}
\label{PPCA}
\begin{eqnarray}
&&\hspace{-0.4in}a_i(t + 1) \in \argmin_{j \in \mathcal{N}} c_i((j,\mathbf{a}(t)_{-i})),\\
&&\hspace{-0.55in}\mbox{and with } \mathbf{a}(t+1) = (a_i(t +
1),\mathbf{a}(t)_{-i}), \nonumber\\
&&\hspace{-0.4in}p_l(t + 1)= c_l(\mathbf{a}(t+1)),\nonumber \\
&& \ \ \ \ \ \ \forall l \in \mathcal{M}_{a_i(t)}(\mathbf{a}(t))
\cup \mathcal{M}_{a_i(t+1)}(\mathbf{a}(t+1)).
\end{eqnarray}
\end{subequations}
\begin{remark}
A mobile $i$ should not choose a BS if the device renders the
corresponding power control subproblem infeasible. The situation is
characterized by $\sum_{l \in \mathcal{M}_j(\mathbf{a})}\beta_l \geq
1$, and~\eqref{cost-fun} appropriately yields infinite cost for the
mobile. Even if the algorithm starts with an infeasible association,
selfish moves of players eventually lead to a feasible one, and
updates remain feasible thereafter.
\end{remark}

Note that while only one mobile updates its association at a time,
all mobiles that perceive a change in load at their BSs update their
powers to optimal values based on the new loads. Simultaneous
association updates are not allowed. In a framework with no
synchronizing agent and with an arbitrarily fine time-scale, it is
unlikely that two mobiles update simultaneously. If two or more BSs
result in the same steady state power, one is chosen at random by
the mobile.


This algorithm is also distributed in nature as the one proposed in
Yates~\cite{ctrltheory-cellular.yates95uplink-power-control}.  BS $j$
broadcasts its total congestion $\sum_{l \in \mathcal{M}_j(\mathbf{a})}\beta_l$. In addition, each mobile $i$ is told its
scaled gains $\frac{h_{ij}}{\sigma^2}$ by each BS $j \in \mathcal{N}$.


\subsection{A Congestion Game Formulation}

To show the convergence properties of the proposed algorithm, we
model the system as a strategic form game. Let the mobiles be the
players and the action set for each player be the possible
associations, i.e, $\mathcal{A}_i = \mathcal{N}$ for all $i \in
\mathcal{M}$.
Define the cost functions of the players to be  $c_i(\mathbf{a})$
for all $i \in \mathcal{M}$. It can be seen that above is a
player-specific singleton weighted congestion game, and belongs to the
subclass of congestion games with multiplicative player-specific
constants described in~\cite{gametheory.mavronicolas-etal07congestion-games-player-specific}.
In the following we refer to it as the strategic form game $(\mathcal{M}, \mathcal{N}, (c_i, i
\in \mathcal{M}))$.



Before analyzing the general game, we consider the following special cases.
\subsubsection{Single Class Traffic}
This is the case where all the mobiles have a common target SINR
requirement $\gamma$; $\beta := \frac{\gamma}{1 + \gamma}$. In this
case,
\[
c_i(\mathbf{a}) = \frac{\sigma^2}{h_{ia_i}} \frac{\beta}{[1 - |\mathcal{M}_{a_i}(\mathbf{a})|\beta ]^+}
\]
and we have a player specific unweighted singleton congestion game.
\subsubsection{Collocated Users}
In  this case, all mobiles are situated close together in a group.
Thus $h_{ij} = h_{j}$ for all $i \in \mathcal{M}, j \in
\mathcal{N}$, and
\[
c_i(\mathbf{a}) = \frac{\sigma^2}{h_{a_i}} \frac{\beta_i}{[1 -  \sum_{l \in \mathcal{M}_{a_i}(\mathbf{a})}\beta_l]^+}.
\]
This yields a player independent weighted singleton congestion game.
\subsubsection{Collocated BS}
Here all BSs are assumed to be situated close together.
Thus $h_{ij} = h_{i}$ for all $i \in \mathcal{M}, j \in \mathcal{N}$, and
\[
c_i(\mathbf{a}) = \frac{\sigma^2}{h_i} \frac{\beta_i}{[1 -  \sum_{l \in \mathcal{M}_{a_i}(\mathbf{a})}\beta_l]^+}.
\]
Now, we get a player specific weighted singleton congestion game.

The following result ensures that MAPC converges in each of these
special cases.
\begin{proposition}
\label{prop:mmpc-conv}
The finite strategic form game $(\mathcal{M},
\mathcal{N}, (c_i, i \in \mathcal{M}))$ is an ordinal potential game
and thus admits the FIP in each of the following  cases.
\begin{enumerate}
\item $\beta_i = \beta$ for all $i \in \mathcal{M}$,
\item $h_{ij} = h_{j}$ for all $i \in \mathcal{M}, j \in \mathcal{N}$,
\item $h_{ij} = h_{i}$ for all $i \in \mathcal{M}, j \in \mathcal{N}$.
\end{enumerate}
\end{proposition}
\begin{IEEEproof}
In each case, we show that the game $(\mathcal{M}, \mathcal{N},
(c_i, i \in \mathcal{M}))$ is better response
equivalent~(Neel~\cite[Chapter~5]{gametheory-cogradio.neel06thesis})
to an ordinal potential game~(by demonstrating an ordinal potential
function for the latter). This implies that, in each case,
 $(\mathcal{M}, \mathcal{N}, (c_i, i \in \mathcal{M}))$ itself is
an ordinal potential game. It is also finite which implies that the FIP
property holds.\\
$1)$ The strategic form game $(\mathcal{M}, \mathcal{N}, (c_i, i \in
\mathcal{M}))$ is better response equivalent to $(\mathcal{M},
\mathcal{N}, (-\frac{1}{c_i}, i \in \mathcal{M}))$. Also note that
\[-\frac{1}{c_i(\mathbf{a})} = -\frac{h_{ia_i}}{\sigma^2} \frac{[1 - |\mathcal{M}_{a_i}(\mathbf{a})|\beta]^+}{\beta}. \]
The function $V_1:\mathcal{N}^{\mathcal{M}} \rightarrow \mathbb{R}$
given by
\[V_1(\mathbf{a}) = -\frac{1}{\sigma^2 \beta} \prod_{l \in \mathcal{M}}h_{la_l} \prod_{k \in \mathcal{N}}\left(\prod_{t = 1}^{|\mathcal{M}_k(\mathbf{a})|}[1 - t\beta]^+\right)\]
satisfies
\begin{eqnarray*}
V_1(j,\mathbf{a}_{-i}) - V_1(\mathbf{a}) = -
\left(\frac{1}{c_i(j,\mathbf{a}_{-i})}-\frac{1}{c_i(\mathbf{a})}\right) \prod_{l \in \mathcal{M} \setminus \{i\}}h_{la_l}\\
\times \prod_{k \in \mathcal{N}}\left(\prod_{t =
1}^{|\mathcal{M}_k(\mathbf{a}) \setminus \{i\}|}[1 -
t\beta]^+\right)
\end{eqnarray*}
for all $i \in \mathcal{M}, j \in \mathcal{N}, \mathbf{a} \in
\mathcal{N}^{\mathcal{M}}$. Thus the game $(\mathcal{M},
\mathcal{N}, (-\frac{1}{c_i}, i \in \mathcal{M}))$ is an ordinal
potential game with a potential function $V_1$.\footnote{This
potential function is similar to those proposed
in~\cite{gametheory-comnet.gairing-etal06routing-player-specific-linear-latency}
for linear cost functions, and
in~\cite{gametheory.mavronicolas-etal07congestion-games-player-specific}
for cost functions
composed of player-specific constants and facility-specific functions.}\\
$2)$ The strategic form game $(\mathcal{M}, \mathcal{N}, (c_i,
i \in \mathcal{M}))$ is better response equivalent to $(\mathcal{M}, \mathcal{N}, (-\frac{\beta_i}{c_i}, i \in
\mathcal{M}))$. Also note that
\[-\frac{\beta_i}{c_i(\mathbf{a})} = -\frac{h_{a_i}[1 - \sum_{l \in \mathcal{M}_{a_i}(\mathbf{a})}\beta_l]^+}{\sigma^2}. \]
For the function $V_2:\mathcal{N}^{\mathcal{M}} \rightarrow
\mathbb{R}$ given by
\[V_2(\mathbf{a}) = -\sum_{i \in \mathcal{M}}\frac{h_{a_i}}{\sigma^2} \beta_i\left(\Big[1 - \sum_{l \in \mathcal{M}_{a_i}}\beta_l\Big]^+ + (1-\beta_i)\right),\]
\[V_2(j,\mathbf{a}_{-i}) - V_2(\mathbf{a}) = -2\left(\frac{\beta_i}{c_i(j,\mathbf{a}_{-i})} - \frac{\beta_i}{c_i(\mathbf{a})}\right)\]
for all $i \in \mathcal{M}, j \in \mathcal{N}, \mathbf{a} \in
\mathcal{N}^{\mathcal{M}}$. Thus $V_2$ is a potential function for
the game $(\mathcal{M}, \mathcal{N}, (-\frac{\beta_i}{c_i}, i \in
\mathcal{M}))$, and so the latter is an ordinal potential game.\\
$3)$ The strategic form game $(\mathcal{M}, \mathcal{N}, (c_i,
i \in \mathcal{M}))$ is better response equivalent to $(\mathcal{M}, \mathcal{N}, (-\frac{\beta_i}{h_i c_i}, i \in
\mathcal{M}))$. Also note that
\[-\frac{\beta_i}{h_i c_i(\mathbf{a})} = -\frac{[1 - \sum_{l \in \mathcal{M}_{a_i}(\mathbf{a})}\beta_l]^+}{\sigma^2}. \]
The function $V_3:\mathcal{N}^{\mathcal{M}} \rightarrow \mathbb{R}$
defined as
\[V_3(\mathbf{a}) = -\sum_{i \in \mathcal{M}}\frac{\beta_i [1 - \sum_{l \in \mathcal{M}_{a_i}}\beta_l]^+}{\sigma^2}\]
satisfies
\[V_3(j,\mathbf{a}_{-i}) - V_3(\mathbf{a}) = -2 \beta_i \left(\frac{\beta_i}{h_i c_i(j,\mathbf{a}_{-i})} - \frac{\beta_i}{h_i c_i(\mathbf{a})}\right)\]
for all $i \in \mathcal{M}, j \in \mathcal{N}, \mathbf{a} \in
\mathcal{N}^{\mathcal{M}}$. Thus the game $(\mathcal{M},
\mathcal{N}, (-\frac{\beta_i}{h_i c_i}, i \in \mathcal{M}))$ is an
ordinal potential game with $V_3$ as a potential function.
\end{IEEEproof}

Now, we focus on the general case. Gairing et
al.~\cite{gametheory-comnet.gairing-etal06routing-player-specific-linear-latency}
show~(via a counter-example with $3$ players) that player-specific
weighted singleton congestion games with linear cost functions are
not necessarily ordinal potential games, and so, need not possess
FIP. This negative result applies to our game also, and convergence
proofs based on potential functions cannot be used. However, it
follows
from~\cite{gametheory.milchtaich96congestion-games-player-specific}
that the strategic form game $(\mathcal{M}, \mathcal{N}, (c_i, i \in
\mathcal{M}))$ admits (i) FIP property if $|\mathcal{N}| = 2$, (ii)
FBRP property if $|\mathcal{M}| = 2$.

Georgiou et al.~\cite{gametheory-comnet.georgiou-etal09network-uncertainty-routing}
establish that player-specific weighted
singleton congestion games with $3$ players and linear cost functions
possess FBRP property.\footnote{On the other hand, Mavronicolas et al.~\cite{gametheory.mavronicolas-etal07congestion-games-player-specific}  demonstrate a best response cycle in a game with $3$ players and costs composed of additive
player-specific constants and facility-specific nondecreasing functions.}
Specifically, they show
in an exhaustive manner that such games do not possess any {\it best response
cycles}.\footnote{There does not seem to be any reason why this technique
cannot be extended to more than $3$ players; though the
number of possibilities in the exhaustive search may become enormous.}
Their result and proof technique extend to
the game $(\mathcal{M}, \mathcal{N}, (c_i, i \in \mathcal{M}))$ even though the cost
functions $c_i$ are not linear. Thus,  the game
$(\mathcal{M}, \mathcal{N}, (c_i, i \in \mathcal{M}))$ can be shown to
possess FBRP if $|\mathcal{M}| = 3$.

\remove{
Mavronicolas et al.~\cite{gametheory.mavronicolas-etal07congestion-games-player-specific}
extend this result to  player-specific weighted singleton congestion games with cost functions
composed of player-specific constants and facility-specific nondecreasing functions. The game
$(\mathcal{M}, \mathcal{N}, (c_i, i \in \mathcal{M}))$ belong to this class, and thus
possesses FBRP if $|\mathcal{M}| = 3$.
}

In case of more than $3$ players, convergence of the best response
dynamics in weighted singleton congestion games with linear cost
functions is an open
problem~\cite{gametheory-comnet.georgiou-etal09network-uncertainty-routing,gametheory.aspnes-etal06open-problems}.
Georgiou et
al.~\cite{gametheory-comnet.georgiou-etal09network-uncertainty-routing}
conjecture that such games always admit at least one NE. Though
functions $c_i$ are not linear, the game $(\mathcal{M}, \mathcal{N},
(c_i, i \in \mathcal{M}))$ is best response equivalent to another
game in which costs are composed of multiplicative player-specific
constants and affine nondecreasing functions. Also, simulations run
on numerous instances of the game suggest that players' updates as
per the best response strategy always converge in a finite number of
steps.
We therefore conjecture that
\begin{conjecture}
\label{conj:fbrp}
 The finite strategic form game $(\mathcal{M},
\mathcal{N}, (c_i, i \in \mathcal{M}))$ admits FBRP and thus
possesses at least one pure strategy Nash equilibrium.
\end{conjecture}

The FBRP property ensures that {\bf MAPC} converges in a finite
number of steps~(see Section~\ref{games-finite}). Consider the
following variants of {\bf MAPC}.
\begin{enumerate}
\item At each $t$, one mobile is randomly chosen to update its association. All mobiles
have strictly positive probabilities of being chosen.
\item At each $t$, each mobile $i$ updates its association with probability
$\epsilon_i \in (0,1)$. There is thus a strictly positive
probability that any subset of mobiles may update their associations
simultaneously. As before, all mobiles update their powers based on
the new loads. This algorithm does not require any coordination
among mobiles~(to ensure one by one updates), and is thus fully
distributed.
\end{enumerate}
The FBRP property of the game $(\mathcal{M}, \mathcal{N}, (c_i, i
\in \mathcal{M}))$ implies that these two algorithms also converge
to a NE with probability $1$ (see
Neel~\cite[Chapter~5]{gametheory-cogradio.neel06thesis}).

\subsection{System Optimality}
\label{sys-opt}

A system optimal power allocation should bring about the lowest
interference environment. This motivates the following definition of
system optimality.

\begin{definition}
For an association profile $\mathbf{a}$, define a system performance
measure $C(\mathbf{a}) = \sum_{i = 1}^{M}{c_i(\mathbf{a})}$ with
$c_i(\mathbf{a})$ defined in~(\ref{cost-fun}). We define an
association profile $\mathbf{a^o}$ to be system optimal if it
minimizes $C(\mathbf{a})$ over all possible associations $\mathbf{a}
\in \times_{i = 1}^M \mathcal{A}_i$.
\end{definition}

Let us now recall the following notion of Pareto
efficiency~\cite[Chapter~5]{commnet-wireless.kumar-etal08wireless-networking}.
\begin{definition}
An association profile $\mathbf{a}$ is said to be Pareto dominated
by another association profile $\mathbf{a'}$ if $c_i(\mathbf{a'})
\leq c_i(\mathbf{a})$ for all $i \in \mathcal{M}$ with
$c_i(\mathbf{a'}) < c_i(\mathbf{a})$ for some $i$. An association
profile $\mathbf{a}$ is said to be Pareto efficient if it is not
Pareto dominated by any other association profile in $\times_{i =
1}^M \mathcal{A}_i$.
\end{definition}

Clearly any association profile that is system optimal is also
Pareto efficient. Thus, if there is a unique Pareto efficient
association profile, it is also the unique system optimal one.
However, unlike the case of single channel networks, joint
association and power control problems in multichannel networks do
not in general admit a unique Pareto efficient power allocation. In
particular, when $|\mathcal{M}| > |\mathcal{N}|$, there cannot be
unique Pareto efficient power allocation.\footnote{Though two Pareto
efficient power allocations can be identical up to a permutation,
e.g., if two mobiles are indifferent  with respect to their SINR
requirements and channel gains to all the BSs.} To see this, define
$\Theta_i$ for any mobile $i$ as the set of best match BSs as
follows
\[\Theta_i :=  \argmin_{j \in \mathcal{N}} \frac{\sigma^2 \gamma_i}{h_{ij}}\]
The system optimal association profile $\mathbf{a^o}$ is clearly
Pareto efficient. Next, two cases are possible.
\begin{enumerate}
\item For all $i$, $a^o_i \in \Theta_i$. Since $|\mathcal{M}| > |\mathcal{N}|$,
there exist two mobiles $i$ and $l$ such that $a^o_{i} = a^o_{l}$.
\item There exists a mobile $i$ such that $a^o_i \notin \Theta_i$.
\end{enumerate}
Consider a mobile $i$ as in Case~1, or as in Case~2. Let
$\mathbf{a}'$ be another profile which is system optimal subject to
$i$ being associated with any of its best match BSs and no other
mobile being associated with that BS. It can be easily checked that
$\mathbf{a}'$ is also Pareto efficient.

As the following example illustrates, {\bf MAPC} may settle at a
Pareto inefficient association profile, and hence may not be system
optimal.
\begin{example}
\label{ex-sys-opt} Consider a network with two BSs, two mobiles, and
a common SINR requirement $\gamma$. The two BSs operate in disjoint
bands. Assume
\begin{eqnarray*}
&& h_{12} < h_{11} < \frac{h_{12}}{(1  - \gamma)} \\
\mbox{and} && h_{21} < h_{22} < \frac{h_{21}}{(1  - \gamma)}.
\end{eqnarray*}
The unique Pareto efficient association is $(a_1 = 1, a_2 =2)$ with
power allocation $(\frac{\sigma^2}{h_{11}}\gamma,
\frac{\sigma^2}{h_{22}}\gamma)$. However, if we start with initial
association  $(a_1 = 2, a_2 = 1)$, {\bf MAPC} will not move forward,
because a unilateral switch requires larger power to meet the target
SINR. Neither mobile will switch to the BS with which it has a
better channel. Hence, $(\frac{\sigma^2}{h_{12}}\gamma,
\frac{\sigma^2}{h_{21}}\gamma)$ is a steady state power vector at
which the algorithm settles; it is Pareto inefficient.
\end{example}

In the following we consider special cases, and investigate whether
the proposed algorithm leads to a system optimal association profile.

\subsubsection{Collocated Mobiles and Single Class Traffic}
Even in this special case, {\bf MAPC} may settle at a Pareto
inefficient NE as shown in the following example.

\begin{example}
Consider a $2$-cell network with $4$ collocated mobiles and $\beta_i =
\beta, i = 1,2,3,4$. Assume that $h_1$ and $h_2$ satisfy
\begin{eqnarray*}
h_1(1 - 3\beta) &=& h_2(1 - 2\beta), \\
h_1(1 - 2\beta) &>& h_2(1 - \beta).
\end{eqnarray*}
The following facts are easily verified. Both the inequalities can
be met simultaneously. The association $(a_1 = a_2 = a_3 = 1, a_4 =
2)$ is a NE from which the algorithm  does not move. This
association is Pareto dominated by $(a_3 = a_4 = 1, a_1 = a_2 = 2)$
which is another NE. Thus {\bf MAPC} may settle at a Pareto
inefficient NE.
\end{example}

Consider now a variant of {\bf MAPC} in which mobile $i = 1 + (t
\mod M)$ updates its association at~$t+1$ if\footnote{The
minimization is with respect to the lexicographical ordering.}
\[
a_i(t) \notin \argmin_{j \in \mathcal{N}} \left(c_i(j,\mathbf{a}(t)_{-i}),\sum_{l \in \mathcal{M}_j(j,\mathbf{a}(t)_{-i})}\beta_l\right),\\
\]
In this case,
\[
a_i(t + 1) \in \argmin_{j \in \mathcal{N}} \left(c_i(j,\mathbf{a}(t)_{-i}),\sum_{l \in \mathcal{M}_j(j,\mathbf{a}(t)_{-i})}\beta_l\right),\\
\]
In words, a mobiles selects a least loaded BS~(counting the tagged
mobile also) among the ones which require transmission with the
least power. We name this variant {\bf MAPC$^{\ast}$}.
\begin{proposition}
\label{prop:mmpc-ast-conv} If the strategic form game $(\mathcal{M},
\mathcal{N}, (c_i, i \in \mathcal{M}))$ contains no best response
cycles, then {\bf MAPC${^\ast}$} converges in a finite number of
steps.
\end{proposition}
\begin{IEEEproof}
Suppose $\mathbf{a^o}$ is the initial association profile. Suppose
$\mathbf{a^1},\cdots,\mathbf{a^k}$ are successive association
profiles generated by {\bf MAPC${^\ast}$}, with $\mathbf{a^1}$
possibly the initial association profile $\mathbf{a^o}$. For an
association profile $\mathbf{a}$, let $\mathbf{m(a)}$ be the
congestion vector with its elements arranged in a decreasing order.
The following two cases are possible.
\begin{enumerate}
\item For some $j \in \{1,\cdots,k-1\}$, the deviating mobile
strictly benefits. Then, $\mathbf{a^k} \neq \mathbf{a^1}$ because
the game $(\mathcal{M}, \mathcal{N}, (c_i, i \in \mathcal{M}))$ does
not contain any best response cycle.
\item The deviating mobiles do not strictly benefit for any $j \in
\{1,\cdots,k-1\}$. Then, it must be that $\mathbf{m(a^1)}, \cdots,
\mathbf{m(a^k)}$ is a lexicographically decreasing sequence. Again,
$\mathbf{a^k} \neq \mathbf{a^1}$.
\end{enumerate}
We thus conclude that iterates generated by {\bf MAPC${^\ast}$}
never contain a cycle. Since there is only a finite number of
distinct association profiles, {\bf MAPC${^\ast}$} must converge in
a finite number of steps.
\end{IEEEproof}

The FBRP property~(Conjecture~\ref{conj:fbrp}) ensures that {\bf
MAPC${^\ast}$} converges in a finite number of steps. We now show
that {\bf MAPC$^{\ast}$} converges to a Pareto efficient NE in the
special case of collocated mobiles and single class traffic.

\begin{proposition}
For the noncooperative game $(\mathcal{M}, \mathcal{N}, (c_i, i \in
\mathcal{M}))$, when $h_{ij} = h_j$ and $\beta_i = \beta$ for all $i
\in \mathcal{M}, j \in \mathcal{N}$, the steady states of {\bf
MAPC$^{\ast}$} are the Pareto efficient NEs of the game.
\end{proposition}
\begin{IEEEproof}
Propositions~\ref{prop:mmpc-conv} and~\ref{prop:mmpc-ast-conv} imply
that MAPC$^{\ast}$ converges in a finite number of steps in this
special case. For any association profile $\mathbf{a}$, let
$m_j(\mathbf{a})$ be the number of mobiles associated with BS $j$.
Let $\mathbf{a}$ be a NE, and $\mathbf{a'}$ be another profile
dominating $\mathbf{a}$. We show that the proposed variant of MAPC
does not settle at $\mathbf{a}$.

We first argue that congestion vectors $\mathbf{m(a)} =
(m_N(\mathbf{a}),\cdots,m_N(\mathbf{a}))$ and $\mathbf{m(a')}$
cannot be identical. Indeed if this is the case, $\mathbf{a'}$ is
obtained by permuting the mobiles' associations in $\mathbf{a}$ in
some way. But then their payoffs undergo the same permutation, which
makes it impossible for all of them to gain.

We define $g_j = \frac{\sigma^2}{h_j}$ and $f(m) = \frac{\beta}{[1 -
m\beta]^+}$. Then a mobile associated with BS $j$ incurs a cost $g_j
f(m_j(\mathbf{a}))$. Further, $\mathbf{a}$ being a NE,
\[g_j f(m_j(\mathbf{a})) \leq g_k f(m_k(\mathbf{a}) + 1)\]
for all $j,k \in \mathcal{N}$. In particular,
\begin{subequations}
\begin{eqnarray}
& m_j(\mathbf{a}) \leq m_k(\mathbf{a}) +
1 &\mbox{if } g_k \leq g_j, \label{load-conditions-1} \\
\mbox{and} & m_j(\mathbf{a}) < m_k(\mathbf{a}) +1 &\mbox{if } g_k <
g_j. \label{load-conditions-2}
\end{eqnarray}
\end{subequations}
Next, we define
\begin{eqnarray*}
\bar{c} &:=& \max_{j \in \mathcal{N}:m_j(\mathbf{a}) > 0} g_j f(m_j(\mathbf{a})),\\
\mbox{and } \mathcal{N}_1 &:=& \argmax_{j \in
\mathcal{N}:m_j(\mathbf{a}) > 0} g_j f(m_j(\mathbf{a}))
\end{eqnarray*}
Under $\mathbf{a}'$ none of the mobiles incurs a cost more than
$\bar{c}$. In particular, those associated with a BS $j \in
\mathcal{N}_1$ under $a'$ must have cost less than $\bar{c}$. This
implies $m_j(\mathbf{a}') \leq m_j(\mathbf{a})$ for all $j \in
\mathcal{N}_1$. Now suppose that $m_j(\mathbf{a}') =
m_j(\mathbf{a})$ for all $j \in \mathcal{N}_1$, and
$m_k(\mathbf{a}')
> m_k(\mathbf{a})$ for a $k \in \mathcal{N} \setminus \mathcal{N}_1$.
Then,
\[g_k f(m_k(\mathbf{a}')) \geq g_k f(m_k(\mathbf{a})+1) \geq g_j
f(m_j(\mathbf{a}))\]
 for any $j \in \mathcal{N}_1$. The last inequality holds because $\mathbf{a}$ is an NE.
Thus we have that
\[g_k f(m_k(\mathbf{a}')) \geq \bar{c},\]
and hence there are more mobiles incurring costs greater than or
equal to $\bar{c}$ under $\mathbf{a}'$ than under $\mathbf{a}$. This
contradicts the hypothesis that $\mathbf{a'}$ Pareto dominates
$\mathbf{a}$. Thus there must be BSs $j \in \mathcal{N}_1, k \in
\mathcal{N} \setminus \mathcal{N}_1$ with $m_j(\mathbf{a}') <
m_j(\mathbf{a})$ and $m_k(\mathbf{a}') > m_k(\mathbf{a})$ which is
same as $m_k(\mathbf{a}') \geq m_k(\mathbf{a}) +1$. Again,
$\mathbf{a}$ being an NE,
\[g_k f(m_k(\mathbf{a}')) \geq \bar{c}.\]
But the hypothesis that $\mathbf{a'}$ Pareto dominates $\mathbf{a}$
implies that
\[
g_k f(m_k(\mathbf{a}')) \leq \bar{c}.
\]
Thus $k$ must belong to the set
\[\mathcal{N}_2 := \{k \in \mathcal{N} \setminus \mathcal{N}_1: g_k f(m_k(\mathbf{a}) + 1) = \bar{c}\}\]
Moreover, $m_k(\mathbf{a}') = m_k(\mathbf{a}) +1$.

Now, we claim that there exist BSs $j \in \mathcal{N}_1$ and $k \in
\mathcal{N}_2$ such that $g_j < g_k$. Assume this claim holds. Then,
\[g_k f(m_k(\mathbf{a}) + 1) = \bar{c} = g_j f(m_j(\mathbf{a}))\]
implies that $m_j(\mathbf{a}) > m_k(\mathbf{a}) + 1$. Thus, under
the proposed algorithm, one of the mobiles associated with BS $j$
moves to BS $k$, i.e., the algorithm does not settle at
$\mathbf{a}$.

We prove the claim via contradiction. Suppose $g_j \geq g_k$ for all
$j \in \mathcal{N}_1, k \in \mathcal{N}_2$. Obtaining $\mathbf{a}'$
from $\mathbf{a}$ may involve three types of load transfers.
\begin{enumerate}
\item One mobile moves from a BS $j \in \mathcal{N}_1$ to a BS $k \in
\mathcal{N}_2$ such that $g_j = g_k$. By the definition of
$\mathcal{N}_2$, such moves only permute the overall cost profile,
and by themselves cannot lead to $\mathbf{a}'$.

\item One mobile moves from a BS $j \in \mathcal{N}_1$ to a BS $k \in
\mathcal{N}_2$ such that $g_j > g_k$. Then, the cost reduces for
$m_j(\mathbf{a}) - 1$ mobiles that are still with BS $j$, but
increases to $\bar{c}$ for $m_k(\mathbf{a}) > m_j(\mathbf{a}) - 1$
mobiles~(see~\eqref{load-conditions-1}). Such moves also cannot lead
to the association profile $\mathbf{a}'$.

\item $n > 1$ mobiles move from a BS $j \in \mathcal{N}_1$ to BSs $k_1,\dots,k_n \in
\mathcal{N}_2$~(they have to move to different BSs, again by the
definition of $\mathcal{N}_2$). Now, the cost reduces for
$m_j(\mathbf{a}) - n$ mobiles, but increases to $\bar{c}$ for
\[\sum_{l=1}^n m_{k_l}(\mathbf{a}) \geq n(m_j(\mathbf{a}) - 1) > m_j(\mathbf{a}) - n \ \ \ \mbox{(see~\eqref{load-conditions-2})}\]
mobiles. Such moves also cannot lead to the association profile
$\mathbf{a}'$.
\end{enumerate}
Thus there must be BSs $j  \in \mathcal{N}_1$ and $k \in
\mathcal{N}_2$ such that $g_j < g_k$ as claimed. This completes the
proof of the proposition.
\end{IEEEproof}

However, the obtained Pareto efficient association profile need not
be system optimal. This is demonstrated by Example~\ref{ex-cont-opt}
for the case of a continuum  of mobiles.

\subsubsection{Collocated BSs}
Next, we consider the case where the BSs are collocated, so that
$h_{ij} = h_i$ for all $i \in \mathcal{M},j \in \mathcal{N}$.
Mobiles may have different target SINR requirements. For any
association profile $\mathbf{a}$, define its {\it support}
$\mathcal{S}_{\mathbf{a}}$ to be the set $\{j \in \mathcal{N}: a_l =
j \mbox{ for some } l \in \mathcal{M}\}$. We say that  $\mathbf{a}$
has {\it full support} if $\mathcal{S}_{\mathbf{a}} = \mathcal{N}$.
\begin{lemma}
\label{NEFSPE}
In the game $(\mathcal{M}, \mathcal{N}, (c_i, i \in \mathcal{M}))$, when $h_{ij} = h_i$ for all
$i \in \mathcal{M}, j \in \mathcal{N}$, any association profile with full support is
Pareto efficient.
\end{lemma}
\begin{IEEEproof}
Consider $\mathbf{a}$, an association profile with full support. Let
there be another association profile $\mathbf{a'}$ that dominates
$\mathbf{a}$.  Assume that $\mathbf{m(a)}$ and  $\mathbf{m(a')}$ are
the congestion vectors corresponding to $\mathbf{a}$ and
$\mathbf{a'}$ respectively. Identify $j_1, \dots, j_N \in
\mathcal{N}$ and $k_1, \dots, k_N \in \mathcal{N}$ such that
$m_{j_1}(\mathbf{a}) \leq \dots \leq m_{j_N}(\mathbf{a})$ and
$m_{k_1}(\mathbf{a'}) \leq \dots \leq m_{k_N}(\mathbf{a'})$. Then we
can show by induction that for every $l = 1, \dots, N$ we must have
$m_{k_l}(\mathbf{a'}) \leq m_{j_l}(\mathbf{a})$. Since $\mathbf{a'}$
dominates $\mathbf{a}$, for a mobile $i$ with $a_i= j_1$ and $a'_i =
k_r$, its costs under $\mathbf{a}$ and $\mathbf{a'}$ must satisfy
\[
\frac{\sigma^2}{h_i} \frac{\beta_i}{[1 - m_{j_1}(\mathbf{a})]^+}
\geq \frac{\sigma^2}{h_i} \frac{\beta_i}{[1 -
m_{k_r}(\mathbf{a'})]^+}.
\]
But $m_{k_r}(\mathbf{a'}) \geq m_{k_1}(\mathbf{a'})$, implying that
\[
\frac{\sigma^2}{h_i} \frac{\beta_i}{[1 - m_{j_1}(\mathbf{a})]^+}
\geq \frac{\sigma^2}{h_i} \frac{\beta_i}{[1 -
m_{k_1}(\mathbf{a'})]^+}.
\]
Moreover, our feasibility assumption ensures that both the sides in
the above inequality are finite. Clearly, $m_{k_1}(\mathbf{a'}) \leq
m_{j_1}(\mathbf{a})$. Assume $m_{k_{l-1}}(\mathbf{a'}) \leq
m_{j_{l-1}}(\mathbf{a})$. Indeed, if $m_{k_l}(\mathbf{a'}) $ were
greater than $m_{j_l}(\mathbf{a})$, then in order to keep their
costs at most as per $\mathbf{a}$, all mobiles associated with the
BSs $j_1,\dots,j_l$ as per $\mathbf{a}$ must be associated with the
BSs $k_1,\dots,k_{l-1}$ in $\mathbf{a'}$. But by the full support
hypothesis $m_{j_l}(\mathbf{a})
> 0$, which when put together with the induction hypothesis yields  $\sum_{r = 1}^l
m_{j_r}(\mathbf{a}) > \sum_{r = 1}^{l-1}m_{k_r}(\mathbf{a'})$. So
all of the mobiles associated with BSs $j_1,\dots,j_l$ under
$\mathbf{a}$ cannot be associated with BSs $k_1,\dots,k_{l-1}$ under
$\mathbf{a'}$. Thus the induction step is proved. Next, since
$\sum_{l = 1}^N m_{j_l}(\mathbf{a}) = \sum_{l = 1}^N
m_{k_l}(\mathbf{a'})$, we must have $m_{k_l}(\mathbf{a'}) =
m_{j_l}(\mathbf{a})$ for all $l = 1, \dots, N$. But under this
condition, all mobiles have the same cost under $\mathbf{a}$ and
$\mathbf{a'}$, contradicting Pareto domination.
\end{IEEEproof}

\begin{proposition}
All the Nash equilibria in the game $(\mathcal{M}, \mathcal{N}, (c_i, i \in \mathcal{M}))$, when $h_{ij} = h_i$ for all
$i \in \mathcal{M}, j \in \mathcal{N}$, are Pareto efficient.
\end{proposition}
\begin{IEEEproof}
Let $\mathbf{a^{\ast}}$ be a Nash equilibrium. The following are the
two possible scenarios.
\begin{enumerate}
\item {\it $\mathbf{a^{\ast}}$ does not have full support:} We must have $|\mathcal{M}| \leq
|\mathcal{N}|$. Indeed, if $|\mathcal{M}| > |\mathcal{N}|$ and
$\mathbf{a^{\ast}}$ does not have full support, then there must be
mobiles $i$ and $l$ with $a^{\ast}_i = a^{\ast}_l$ and a BS $j$ with
$\mathcal{M}_j(\mathbf{a^{\ast}}) = \emptyset$. Clearly, mobile $i$
benefits by moving to BS $j$. This contradicts the fact that
$\mathbf{a^{\ast}}$ is a NE.

Next, $i \neq l$ implies $a^{\ast}_i \neq a^{\ast}_l$ for the same
reason as explained above. Since all BSs have the same channel gain
to a mobile, $\mathbf{a^{\ast}}$ is Pareto efficient.
\item {\it $\mathbf{a^{\ast}}$ has full support:} Lemma~\ref{NEFSPE} implies that $\mathbf{a^{\ast}}$ is Pareto
efficient
\end{enumerate}
\end{IEEEproof}

However, a NE need not be system optimal if
traffic is not of single class or mobiles are not collocated as shown in the following examples.

\begin{example}
\label{ex-col-mobiles-BSs} Consider a 2-cell network with collocated
BSs. The 4 mobiles are symmetrically located with respect to the 2
collocated BSs. Thus $h_j = h,j = 1,2$. For this example assume
$\beta_i = 2\beta,i = 1,2$, and $\beta_i = 3\beta,i = 3,4$. It can
be seen that $(a_1 = a_2 = 1, a_3 = a_4  = 2)$ is a NE which is
Pareto efficient. But its social cost is higher than that of $(a_1 =
a_3 = 1, a_2 = a_4  = 2)$ which is another NE.
\end{example}

\begin{example}
\label{ex-col-BSs-single-class} Consider a 2-cell network with 5
mobiles. The 2 BSs are collocated. Further, $h_i = ih$ and $\beta_i
= \beta, i = 1,2,3,4$ where $\frac{1}{4} < \beta < \frac{1}{3}$. Any
profile in which two mobiles associate with one BS, and the
remaining three with another is a NE. On the other hand, $(a_1 = a_2
= 1, a_3 = a_4 = a_5 = 2)$ and $(a_1 = a_2 = 2, a_3 = a_4  = a_5 =
1)$ are the only socially optimal NEs.
\end{example}

\subsubsection{Collocated Mobiles, Symmetrically Placed BSs, and Single Class Traffic}
In the special case when all the mobiles are collocated and all the
BSs are symmetrically placed with respect to the collocated mobiles,
we have the following result.
\begin{proposition}
All the NEs in the game $(\mathcal{M}, \mathcal{N}, (c_i, i \in
\mathcal{M}))$, with $h_{ij} = h$ for all $i \in \mathcal{M}, j \in
\mathcal{N}$, are system optimal.
\end{proposition}
\begin{IEEEproof}
The mobiles as well as BSs are indistinguishable in this game. At a
NE, let $m_j$ be the number of mobiles associated with BS $j$. We
first prove that at any NE, the vector of mobiles' costs is unique
up to permutations. To prove this, it suffices to prove that the
vector $\mathbf{m} = (m_j, j \in \mathcal{N})$ for a NE is unique up
to permutations. As $\mathbf{m}$ yields a NE, the following must
hold for all $j,k \in \mathcal{N}$:
\begin{eqnarray}
\frac{\sigma^2}{h} \frac{\beta}{1 - m_j\beta} &\leq& \frac{\sigma^2}{h} \frac{\beta}{1 - m_k\beta - \beta} \nonumber \\
\mbox{or } m_j &\leq& m_k + 1. \label{eq-sym-collocated}
\end{eqnarray}
Define $n = \lfloor \frac{M}{N} \rfloor$ and $l = M - nN$.
From~(\ref{eq-sym-collocated}) we see that $\mathbf{m}$ given by
$m_j = n + 1, j = 1,\dots,l,m_j = n,j = l+1,\dots,N$ characterizes
one of the NEs; other NEs are permutations of this vector, and $m$
is unique up to permutations. We now show that $\mathbf{m}$ is a
system optimal congestion vector, and the system optimality of all
other NEs follows. To do this observe that
\[C(\mathbf{a}) = \frac{\sigma^2}{h}\sum_{i \in \mathcal{M}}\frac{\beta}{1 - m_{a_i}\beta} = \frac{\sigma^2}{h}\sum_{j \in \mathcal{N}}\frac{m_j\beta}{1 - m_j\beta}\]
is a {\it Schur-convex} function in $(m_1,\dots,m_N)$ because
$\frac{x}{1-x}$ is a convex function. This implies that the minimum
value is attained at a vector which is as close to uniform as
possible, i.e., a vector that is {\it majorized} by any other
vector~(Marshall \&
Olkin~\cite{ctrltheory-optimization.marshall-olkin79majorization}).\footnote{The
condition $h_{ij} = h_i$ for all $j \in \mathcal{N}$ is used to
deduce that NE profiles are majorized by any non NE profile; the
condition $h_{ij} = h_j$  for all $i \in \mathcal{M}$ is used to
deduce Schur-convexity of $C(a)$.} All such vectors are permutations
of $\mathbf{m}$~(Alternatively, if there exist BSs $j$ and $k$ such
that $m_j \geq m_k + 2$, moving a mobile from BS $j$ to BS $k$
results in a strictly lower cost). This concludes the proof.
\end{IEEEproof}

\section{Continuum of  Mobiles}

\label{cont-users} In this section, we consider a nonatomic version
of the system in Section~\ref{sys-model}. Such a model is of
interest for two reasons. First, for many of the fixed QoS traffic
classes (e.g., voice), the target SINR requirements in CDMA cellular
systems are very small. In a typical IS 95 CDMA system with system
bandwidth 1.25~MHz, chip rate 1.2288~Mcps, data rate 9.6~Kbps, and
target $\frac{E_b}{N_0} =$ 6~dB, the target SINR turns out to be
-15~dB, i.e., $\frac{1}{32}$~(Kumar
et~al.~\cite[Chapter~5]{commnet-wireless.kumar-etal08wireless-networking}).
If we assume that at any time the number of mobiles associated with
a BS is large, it is reasonable to say that an incoming mobile or an
outgoing mobile has a negligible effect on the congestion. Secondly,
we have seen that our proposed algorithm may end up with inefficient
associations. There is extensive work on toll mechanisms that induce
system optimality in networks with a continuum of users. The
analysis of toll-mechanisms (or pricing) on a multichannel multicell
network with a continuum of mobiles can be expected to shed light on
the existence and properties of pricing mechanisms for networks with
discrete mobiles.

\subsection{System Model}
 Let $\mathcal{M} = \cup_{l=1}^{L}\mathcal{M}_l$ be an infinite set of $\mathcal{L} = \{1,\dots,L\}$ classes of nonatomic mobiles. By nonatomic mobiles, we mean that the effect of a single mobile at a BS is infinitesimal. The population of class $l$ mobiles has ``mass'' $M_{l}$.  
All the mobiles in a class are collocated and require equal minimum
SINR. In particular, all such mobiles have same power gains to any
of the BSs~(gains from a mobile to different BSs can be different).
Assume  $\mathcal{N}$ to be the finite set of  BSs.  As before,
$\sigma$ denotes the common standard deviation of receiver noise at
all BSs. Let $\gamma_l$ be the common minimum required SINR density
for class $l$ mobiles, and $h_{lj}$ be the power gain between a
class $l$ mobile and BS $j$. An association profile $a$ is a
measurable function $a:\mathcal{M} \rightarrow \mathcal{N}$. Any
association $a$ leads to a congestion profile $\mathbf{m}(a) =
(m_{lj}(a), l \in \mathcal{L}, j \in \mathcal{N})$, $m_{lj}(a)$
being the mass of class $l$ mobiles associated with BS $j$. Let
$\mathcal{M}^{\ast}$ denote the set of all such congestion profiles.

Under an association profile $a$ and a power density allocation
$p:\mathcal{M} \rightarrow \mathbb{R}_+$, the SINR density for $x
\in \mathcal{M}_l,l \in \mathcal{L}$ is
\begin{eqnarray*}
&& \frac{h_{la(x)}p(x)}{\sum_{l=1}^{L}\int_{\mathcal{M}_l}1_{a}(x,z)h_{la(z)}p(z)dz + \sigma^2},  \\
\mbox{where} &&  1_{a}(x,z) = \left\{ \begin{array}{ll}
                              1, \mbox{ if } a(x) = a(z) \\
                              0, \mbox{ otherwise } \end{array} \right. \end{eqnarray*}

Our definition of a ``class'' makes all the mobiles in a class alike, and so,
congestion profiles are sufficient to characterize the system.
In the sequel, we just use $m_{lj}$ for $m_{lj}(a)$ for convenience. The dependence on $a$ is understood.

Consider again the subproblem of power control with a fixed congestion profile $\mathbf{m}$.
The following result is analogous to Proposition~\ref{dis-steady-pow}, and is shown in Appendix~\ref{nonatomic-pc}.

\begin{proposition}
\label{cont-steady-pow}
\begin{enumerate}
\item The power control subproblem of BS $j$ is feasible iff $\sum_{l \in \mathcal{L}}m_{lj}\gamma_l < 1$.
\item If the power control subproblem of BS $j$ is feasible,
there exists a unique Pareto efficient power density $p$ given by
\[p(x) = \frac{\sigma^2}{h_{lj}} \frac{\gamma_l}{1 - \sum_{l \in \mathcal{L}}m_{lj}\gamma_l},\]
$\forall x \in \mathcal{M}_l$ such that $a(x) = j, l \in \mathcal{L}$, where $a$ is the
underlying association profile.
\end{enumerate}
\end{proposition}

An evolutionary dynamics can be proposed to address the combined association and power control problem.
To this end, we define functions $c_{lj}:\mathcal{M}^{\ast}
\rightarrow \mathbb{R}_+$, where $c_{lj}(\mathbf{m})$ denotes the
minimum power density for class $l$ mobiles associated with BS $j$
under congestion profile $\mathbf{m}$, as
\begin{equation*}
c_{lj}(\mathbf{m}) =  \frac{\gamma_l \sigma^2}{h_{lj}[1 - \sum_{l
\in \mathcal{L}}m_{lj}\gamma_l]^+}.
 \end{equation*}
For notational convenience, define
\begin{eqnarray*}
&& g_{lj} = \frac{\gamma_l \sigma^2}{h_{lj}}, \\
&& m_j = \sum_{l = 1}^{L} \gamma_l m_{lj}, \forall j \in \mathcal{N}\\
\mbox{and} && c(z) = \left\{ \begin{array}{ll}
                     \frac{1}{1 - z}, \mbox{ if } z < 1 \\
                     \infty,  \mbox{ if } z \geq 1. \end{array} \right. \end{eqnarray*}
We then have
\begin{equation}
\label{cost-fun-nonatomic} c_{lj}(\mathbf{m}) = g_{lj}c(m_j).
\end{equation}
Again we assume that the system is feasible, i.e., there exists a
feasible assignment, as done in Section~\ref{atomic-assoc}. This
boils down to assuming $\sum_{l \in \mathcal{L}} \gamma_l M_l < N$
in the case of nonatomic mobiles. Now, structures of the cost
functions allow us to restrict attention to the region where $m_j <
1, \forall j \in \mathcal{N}$; if $m_j \geq 1$ for a $j \in
\mathcal{N}$, all the mobiles associated with $j$ incur infinite
cost.


\subsection{A Congestion Game Formulation}
We model the problem as a nonatomic congestion game.
The continuum of mobiles constitute the population, and $\mathcal{N}$ denotes the
common action set for players of all the classes.
Class $l$ players are accompanied by cost functions $c_{lj}(\mathbf{m}), j \in \mathcal{N}$.
In the following, we refer to it as the
game $(\mathcal{M}, \mathcal{N}, (c_{lj}, l \in \mathcal{L}, j \in \mathcal{N}))$.




\begin{proposition}
\label{nonatomic-potential} The nonatomic game $(\mathcal{M},
\mathcal{N}, (c_{lj}, l \in \mathcal{L}, j \in \mathcal{N}))$ is a
potential game. Furthermore, it admits at least one NE, and the set
of NEs coincides with the set of minimizers of the potential
function.
\end{proposition}
\begin{IEEEproof}
In the region $\{\mathbf{m} : m_j < 1, \forall j \in \mathcal{N}\}$,
the function $V: \mathcal{M}^{\ast} \rightarrow \mathbb{R} \cup
\{+\infty\}$ defined as
\begin{equation}
V(\mathbf{m}) :=  \sum_{j \in \mathcal{N}}\left(\sum_{l \in
\mathcal{L}}\gamma_l m_{lj}\log g_{lj} + \int_{0}^{m_j} \log c(x)dx
\right), \label{nash-nonatom-obj}
\end{equation}
is a $\mathbf{C}^1$ function with
\[\frac{\partial V(\mathbf{m})}{\partial m_{lj}} = \gamma_l \log g_{lj} + \gamma_l \log c(m_j) = \gamma_l \log c_{lj}(\mathbf{m})\]
for all $l \in \mathcal{L}, j \in \mathcal{N},  \mathbf{m} \in
\mathcal{M}^{\ast}$. Thus the nonatomic game $(\mathcal{M},
\mathcal{N}, (\gamma_l \log c_{lj}, l \in \mathcal{L}, j \in
\mathcal{N}))$ is a potential game with $V(\mathbf{m})$ as a
potential function~(see Definition~\ref{cont-potential}). Note that
the strategic form game $(\mathcal{M}, \mathcal{N}, (c_{lj}, l \in
\mathcal{L}, j \in \mathcal{N}))$ is better response equivalent to
$(\mathcal{M}, \mathcal{N}, (\gamma_l \log c_{lj}, l \in
\mathcal{L}, j \in \mathcal{N}))$. Thus the former is also a
potential game with the same potential function  $V(\mathbf{m})$.

Now consider the following optimization problem
\begin{subequations}
\begin{eqnarray}
\mbox{Minimize} && V(\mathbf{m}) \nonumber \\
\mbox{subject to} && \sum_{j \in \mathcal{N}} m_{lj} = M_l, \ l \in \mathcal{L} \label{nash-nonatom-cond1}\\
                  && m_{lj} \geq 0, \ l \in \mathcal{L}, j \in \mathcal{N}.\label{nash-nonatom-cond2}
\end{eqnarray}
\end{subequations}
All the conditions are self-explanatory. \remove{ The first
claimfollows from
Sandholm~\cite[Section~2]{gametheory.sandholm01potential-games-continuous-player-sets},
which also shows that the objective
function~(\ref{nash-nonatom-obj}) is a potential function for the
population game $(\mathcal{M}, \mathcal{N}, (c_{lj}, 1 \leq l \leq
L, j \in \mathcal{N}))$. }
 Observe that
\[\frac{\partial^2 V(\mathbf{m})}{\partial m_{ik} m_{lj}} = \left\{ \begin{array}{ll}
                                                            \gamma_i \gamma_l c(m_j) &\mbox{if } k = j, \\
                                                             0 &\mbox{otherwise}.
                                                               \end{array} \right.  \]
Thus, with an appropriate ordering of the components of $m$, the
hessian of $V(\mathbf{m})$ is given by
\[
\nabla^2 V(\mathbf{m}) = \begin{bmatrix}
 c(m_1)D &  0  & \ldots & 0\\
0  &  c(m_2)D & \ldots & 0\\
\vdots & \vdots & \ddots & \vdots \\
0  &   0 &\ldots & c(m_N)D
\end{bmatrix},
\]
where
\begin{subequations}
\begin{eqnarray}
D &:=& \Gamma \Gamma^T, \label{defn-D1}\\
\mbox{and } \Gamma &:=& [\gamma_1,\cdots,\gamma_L]^T.
\label{defn-D2}
\end{eqnarray}
\end{subequations}
Clearly, $D$, and hence $\nabla^2 V(\mathbf{m})$ is a positive
semi-definite matrix. Thus, $V(\mathbf{m})$ is a convex function of
$\mathbf{m}$. Since we are minimizing a convex objective function
subject to linear constraints, there exists at least one minimizer,
and all minima are global minima. Also, Kuhn-Tucker first order
conditions are necessary and sufficient~\cite[Section~5.5.3]{BV09}.
Combining this with the fact that NEs are the profiles which satisfy
the Kuhn-Tucker first order conditions for a minimizer of the
potential function~(see Section~\ref{pop-games}), we see that the
set of NEs coincides with the set of minimizers of the potential
function.
\end{IEEEproof}

\begin{remark}
The assertion in the above proposition does not hold for general
population games. While all local maximizers of potential function
are equilibria, not all equilibria maximize potential~(even locally)
in
general~\cite[Section~3]{gametheory.sandholm01potential-games-continuous-player-sets}.
This is unlike finite player potential games where only equilibria
are the local maximizers of potential functions.
\end{remark}

Furthermore, NEs have the following
property~(see~\cite[Proposition~3.3]{gametheory.milchtaich00uniqueness-equilibrium-large-crowding-games}).
\begin{proposition}
\label{cost-unique}
The congestion at a BS is constant across all the NEs of the
game $(\mathcal{M}, \mathcal{N}, (c_{lj}, l \in \mathcal{L}, j \in \mathcal{N}))$.
Consequently, the cost density for a class is also constant across all the NEs.
\end{proposition}
\begin{remark}
\label{re-cong-nash} At NEs, the congestions~(at BSs) by class,
$m_{lj}$, are not unique because the objective
function~(\ref{nash-nonatom-obj}) is not {\it strictly} convex with
respect to this set of variables.
\end{remark}

\subsection{System Optimality}
Analogous to the definition in Section~\ref{sys-opt}, we define the
system performance measure
\begin{equation}
\label{obj-function-cont}
 C(\mathbf{m}) := \sum_{j \in \mathcal{N}} \sum_{l =
 1}^{L}m_{lj}g_{lj}c(m_j).
\end{equation}
A congestion profile $\mathbf{m^{\ast}} \in \mathcal{M}^{\ast}$ is
said to be system optimal if it minimizes $C(\mathbf{m})$ over all
possible profiles $\mathbf{m} \in  \mathcal{M}^{\ast}$.

In contrast with the discrete mobiles case where equilibria need not be Pareto efficient~(see Example~\ref{ex-sys-opt}),
we have the following result for the nonatomic case.

\begin{proposition}
All NEs of the  nonatomic game $(\mathcal{M}, \mathcal{N}, (c_{lj}, l \in \mathcal{L}, j \in \mathcal{N}))$ are Pareto efficient.
\end{proposition}
\begin{IEEEproof}
Let $\mathbf{m}$ be a NE congestion profile. Under a NE, the cost
densities for the mobiles of the same class are equal, irrespective
of their associations~(see Remark~\ref{pop-NE}).  Thus, it is
sufficient to prove that there does not exist another congestion
profile $\mathbf{m'}$ such that for every class $l$, and for all BSs
$j,k$, with $m_{lj} > 0, m'_{lk} > 0$,
\begin{equation}
\label{pareto-nonatom}
c_{lk}(\mathbf{m'}) \leq c_{lj}(\mathbf{m}),
\end{equation}
and strict inequality holds for some such $l,j$ and $k$.
Assume that such an $\mathbf{m'}$ exists. Then,
\[g_{lk} c(m'_k) < g_{lj} c(m_j) \leq g_{lk} c(m_k)\]
where the last inequality follows because $\mathbf{m}$ is a NE and
$m_{lj} > 0$. This yields $m'_k < m_k$. This further implies that
there is a BS $s$ such that $m'_s > m_s$, and a class $t$ such that
$m'_{ts} > m_{ts}$. By the strictly increasing property of $c$, we
have
\[g_{ts} c(m'_s) > g_{ts} c(m_s) \geq g_{tr}c(m_r) \]
for a BS $r$ such that $m_{tr} > 0$. Such a BS of course exists and
the latter inequality follows because $\mathbf{m}$ is a NE. The two
inequalities imply $c_{ts}(\mathbf{m'}) > c_{tr}(\mathbf{m})$, and
so the tuple $t,r,s$ violates~(\ref{pareto-nonatom}). Thus the
assumption that  $\mathbf{m'}$ Pareto dominates $\mathbf{m}$ is
incorrect. This completes the proof.
\end{IEEEproof}

We show that the NEs are system optimal if all the mobiles are
collocated, and all the BSs are symmetrically placed around them.
\begin{proposition}
All NEs in the nonatomic game $(\mathcal{M}, \mathcal{N}, (c_{lj}, l
\in \mathcal{L}, j \in \mathcal{N}))$, with $h_{lj} = h$ for all $l
\in \mathcal{L},j \in \mathcal{N}$, are system optimal.
\end{proposition}
\begin{IEEEproof}
In the case of collocated base stations
\begin{eqnarray*}
 C(\mathbf{m}) &=&  \frac{\sigma^2}{h} \sum_{j \in \mathcal{N}} \sum_{l =  1}^{L} \gamma_l m_{lj}c(m_j) \\
               &=&  \frac{\sigma^2}{h} \sum_{j \in \mathcal{N}} m_j c(m_j)
\end{eqnarray*}
For the reason described earlier, we restrict attention to the
region where $m_j < 1, \forall j \in \mathcal{N}$. In this region,
\[\frac{d}{dm_j}m_jc(m_j) = \frac{1}{(1 - m_j)^2},\]
and so $m_j c(m_j)$ is a convex function of $m_j$. Thus
$C(\mathbf{m})$ is a Schur-convex function of $(m_j, 1 \leq j \leq
N)$, and is minimized at any $\mathbf{m^{\ast}}$ with
\[m^{\ast}_j = \frac{1}{N}\sum_{l \in \mathcal{L}}\gamma_l M_l\]
for all $j \in \mathcal{N}$. When $h_{lj} = h$ for all $l \in
\mathcal{L},j \in \mathcal{N}$, any congestion profile with equal
congestion at all the BSs is a NE. Thus, the system optimal profile
$\mathbf{m^{\ast}}$ is a NE. Since all the NEs incur equal cost~(see
Proposition~\ref{cost-unique}), all NEs are system optimal.
\end{IEEEproof}

However, NEs need not be system optimal if BSs are not collocated,
or mobile are not collocated. We illustrate these facts through the
following examples.
\begin{example}
\label{ex-cont-opt2} Consider an infinite set $\mathcal{M}$ of
nonatomic mobiles belonging to two classes; class~1 and class~2
mobiles have masses $M_1$ and $3M_1$ respectively. Assume common
minimum SINR density requirement $\gamma$, and let $3M_1\gamma < 1$.
Let there be two collocated BSs. Let the power gain between a
class~$l$ mobile and a BS be $h_l$, $h_1 < \frac{h_2}{3}$. A
congestion profile is a NE if and only if it assigns equal load to
both the BSs. Thus, the total cost incurred at NE
\[C^{\ast} = \frac{\gamma M_1 \sigma^2}{h_1(1 - 2 \gamma M_1)} + \frac{3\gamma M_1 \sigma^2}{h_2(1 - 2\gamma M_1)}.\]
Next, consider a profile in which class~1 mobiles associate with
BS~1 and class~2 mobiles associate with BS~2. The cost incurred now
is
\[C = \frac{\gamma M_1 \sigma^2}{h_1(1 - \gamma M_1)} + \frac{3\gamma M_1 \sigma^2}{h_2(1 - 3\gamma M_1)}.\]
It can be easily checked that $C < C^{\ast}$ if
\[M_1 < \frac{1}{\gamma}\frac{h_2 / 3 - h_1}{h_2 - h_1}.\]
\end{example}

\begin{example}
\label{ex-cont-opt} Consider an infinite set $\mathcal{M}$ of
nonatomic mobiles all belonging to same class; $M := |\mathcal{M}|$.
Assume common minimum SINR density requirement $\gamma$, and  let
$M\gamma < 1$. Let there be two BSs with $h_j$ the gain to BS $j$,
$j = 1,2$. An NE congestion profile $(\alpha^{\ast}M,
(1-\alpha^{\ast})M)$ is given as
\begin{enumerate}
\item if $\frac{h_1}{h_2} \leq (1 - M\gamma)$, $\alpha^{\ast} = 0$,
\item if $\frac{h_2}{h_1} \leq (1 - M\gamma)$, $\alpha^{\ast} = 1$,
\item otherwise, $\alpha^{\ast}$ satisfies
\begin{eqnarray}
\frac{\gamma \sigma^2}{h_1(1 - \alpha^{\ast}\gamma M)} &=& \frac{\gamma \sigma^2}{h_2(1 - (1-\alpha^{\ast})\gamma M)} \nonumber \\
\mbox{i.e., } \frac{1 - \alpha^{\ast}\gamma M}{1 - (1 -
\alpha^{\ast})\gamma M} &=& \frac{h_2}{h_1} \label{eq-cont-nash}.
\end{eqnarray}
\end{enumerate}
On the other hand, a congestion profile $(\alpha^oM, (1-\alpha^oM)$
will be system optimal if and only if $\alpha^o$ solves the
following optimization problem:
\begin{eqnarray}
&& \hspace{-0.5in} \mbox{Minimize } \ \frac{\alpha \gamma M \sigma^2}{h_1(1 - \alpha \gamma M)} + \frac{(1-\alpha) \gamma M \sigma^2}{h_2(1 - (1-\alpha)\gamma M)} \label{exm-obj-function} \\
&& \hspace{-0.5in} \mbox{subject to } \ 0 \leq \alpha \leq 1. \nonumber
\end{eqnarray}
This is a convex optimization problem, and it is straightforward to show that
\begin{enumerate}
\item if $\sqrt{\frac{h_1}{h_2}} \leq (1 - M\gamma)$, $\alpha^o = 0$,
\item if $\sqrt{\frac{h_2}{h_1}} \leq (1 - M\gamma)$, $\alpha^o = 1$,
\item otherwise, $\alpha^o$ satisfies
\begin{equation}
\frac{1 - \alpha^o \gamma M}{1 - (1 - \alpha^o)\gamma M} =
\sqrt{\frac{h_2}{h_1}} \label{eq-cont-opt}
\end{equation}
\end{enumerate}
Hence, if $\min\{\frac{h_1}{h_2}, \frac{h_2}{h_1}\} > 1 - M\gamma$,
then $\min\{\sqrt{\frac{h_1}{h_2}}, \sqrt{\frac{h_2}{h_1}}\}
> 1 - M\gamma$, and $\alpha^{\ast}$ and  $\alpha^o$ must
satisfy~(\ref{eq-cont-nash}) and~(\ref{eq-cont-opt}) respectively.
In such a case, the NE will be system optimal if and only if $h_1 =
h_2$.
\end{example}

\begin{remark}
Sandholm~\cite{gametheory.sandholm01potential-games-continuous-player-sets} shows that if the cost function for each mobile is a homogeneous function of a certain degree, then all NEs are system optimal. Note that in Example~\ref{ex-cont-opt}, NEs are not system optimal unless $h_1 = h_2$. We remark that the system optimality for the latter case does not follow from Sandholm~\cite{gametheory.sandholm01potential-games-continuous-player-sets} because the cost functions are not homogeneous functions.
\end{remark}

\section{Pricing for System Optimality}
\label{pricing}

\subsection{Continuum of Mobiles}

In this section, we show that there is a toll mechanism that can
induce system optimal associations and power allocations in a
cellular network with multiple classes of mobiles. We also show that
the mechanism can be employed in a distributed fashion.

Define
\begin{equation*}
c'(z) := \left\{ \begin{array}{ll}
                \frac{d}{dz}c(z) = \frac{1}{(1 - z)^2}, &\mbox{ if } z < 1  \\
                \infty,  &\mbox{ if } z \geq 1 \end{array} \right.
\end{equation*}
Consider a congestion profile $\mathbf{m} = (m_{lj}, l \in
\mathcal{L}, j \in \mathcal{N})$. We propose that a class $l$ mobile
joining BS $j$ be levied a toll
\begin{equation}
t_{lj}(\mathbf{m}) = \gamma_l\sum_{i = 1}^{L}m_{ij} g_{ij}c'(m_j).
\label{eq-cont-toll}
\end{equation}
Now, define $\bar{c}_{lj}(\cdot) = c_{lj}(\cdot) + t_{lj}(\cdot),
\forall l \in \mathcal{L}, j \in \mathcal{N}$, and consider the
nonatomic game $(\mathcal{M}, \mathcal{N}, (\bar{c}_{lj}, l \in
\mathcal{L}, j \in \mathcal{N}))$. Players may incur different power
costs ($c_{lj}(\cdot)$) in different NEs of this game. Therefore,
one has to distinguish between the following two cases~(see Fotakis
\&
Spirakis~\cite{gametheory-comnet.fotakis-spirakis07cost-balancing-tolls}).
\begin{enumerate}
\item A toll mechanism is said to {\it weakly enforce} system optimality
 if some NE of the game with tolls is an optimal profile.
\item It is said to {\it strongly enforce} system
optimality if all the NEs of the game with tolls are optimal
profiles.
\end{enumerate}
We show that tolls $t_{lj}(\cdot)$ weakly enforce system optimality
in all cases and strongly enforce it in a special setting.

\begin{proposition}
\label{opt-nonatom} The nonatomic game $(\mathcal{M}, \mathcal{N},
(\bar{c}_{lj}, l \in \mathcal{L}, j \in \mathcal{N}))$ is a
potential game. Furthermore, a congestion profile $\mathbf{m}$ is
system optimal only if it is a NE of this game.
\end{proposition}
\begin{IEEEproof}
Recall the system performance measure~$C(\mathbf{m})$ defined
in~\eqref{obj-function-cont}. Observe that
\[
\frac{\partial C(\mathbf{m})}{\partial m_{lj}} =
\bar{c}_{lj}(\mathbf{m})
\]
for all $l \in \mathcal{L}, j \in \mathcal{N}$, and $\mathbf{m}$ in
$\{\mathbf{m} : m_j < 1, \forall j \in \mathcal{N}\}$. Thus the
nonatomic game $(\mathcal{M}, \mathcal{N}, (\bar{c}_{lj}, l \in
\mathcal{L}, j \in \mathcal{N}))$ is a potential game with
$C(\mathbf{m})$ as a potential function~(see
Definition~\ref{cont-potential}).

Next, observe that system optimal associations and powers are
solutions of the following nonlinear optimization problem:
 \begin{eqnarray*}
\mbox{Minimize} && C(\mathbf{m})\\
\mbox{subject to} &&
\mbox{Conditions~(\ref{nash-nonatom-cond1})~-~(\ref{nash-nonatom-cond2})}.
\end{eqnarray*}
Since all the constraints are linear, any optimizing congestion
profile of $C(\mathbf{m})$ necessarily satisfies the
Karush-Kuhn-Tucker first order conditions~\cite[Chapter~5]{BV09}.
But, any congestion profile satisfying these conditions is a NE of
the game $(\mathcal{M}, \mathcal{N}, (\bar{c}_{lj}, l \in
\mathcal{L}, j \in \mathcal{N}))$. Thus, any system optimal
congestion profile is also a NE of this game.
\end{IEEEproof}

If all the mobiles are collocated, the proposed tolls strongly
enforce system optimality.

\begin{proposition}
All NEs in the nonatomic game $(\mathcal{M}, \mathcal{N},
(\bar{c}_{lj}, l \in \mathcal{L}, j \in \mathcal{N}))$, with $h_{lj}
= h_j$ for all $l \in \mathcal{L},j \in \mathcal{N}$, are system
optimal.
\end{proposition}
\begin{IEEEproof}
It suffices to show that $C(\mathbf{m})$ is a convex function if
$h_{lj} = h_j$ for all $l \in \mathcal{L},j \in \mathcal{N}$. Then,
any congestion profile satisfying Karush-Kuhn-Tucker
conditions~(i.e., any NE) is system optimal.

If $h_{lj} = h_j$ for all $l \in \mathcal{L},j \in \mathcal{N}$,
\[C(\mathbf{m}) = \sum_{j \in \mathcal{N}} \frac{ m_j c(m_j)\sigma^2}{h_j}.\]
Using the observation $c(x) + x c'(x) = c'(x)$, it is easy to see
that
\[
\frac{\partial^2 C(\mathbf{m})}{\partial m_{ik} m_{lj}} = \left\{
\begin{array}{ll}
                                                            \frac{\gamma_i \gamma_l c''(m_j) \sigma^2}{h_j}&\mbox{if } k = j, \\
                                                             0 &\mbox{otherwise},
                                                               \end{array}
                                                               \right.
\]
and
\[\nabla^2 C(\mathbf{m}) = \begin{bmatrix}
 \frac{c''(m_1) \sigma^2}{h_1}D &  0  & \ldots & 0\\
0  &   \frac{c''(m_2) \sigma^2}{h_2}D & \ldots & 0\\
\vdots & \vdots & \ddots & \vdots \\
0  &   0 &\ldots &  \frac{c''(m_N) \sigma^2}{h_N}D
\end{bmatrix},\]
with $D$ given by~\eqref{defn-D1}-\eqref{defn-D2}. It is now obvious
that $\nabla^2 C(\mathbf{m})$ is a positive semi-definite matrix,
and so $C(\mathbf{m})$ is a convex function of $\mathbf{m}$.
\end{IEEEproof}

However, the tolls $t_{lj}(\mathbf{m})$ may fail to strongly enforce
a system optimal congestion profile even if all the BSs collocated,
the mobiles require a constant SINR density $\gamma$, but they are
not collocated. To see this, consider the congestion profile
$\mathbf{m^{\ast}}$ with
\[m^{\ast}_{lj} = \frac{M_l}{N} \ \forall l \in \mathcal{L}, j \in \mathcal{N}.\]
It can be easily checked that, for all
$l \in \mathcal{L}$,
\[\bar{c}_{lj}(\mathbf{m^{\ast}}) = \frac{\gamma \sigma^2}{h_l}c\left(\frac{\gamma M}{N}\right) + \sum_{i = 1}^{L} \frac{\gamma^2 \sigma^2}{h_i} \frac{M_i}{N} c'\left(\frac{\gamma M}{N}\right),\]
which is independent of $j \in \mathcal{N}$. Thus
$\mathbf{m^{\ast}}$ is a NE of the game $(\mathcal{M}, \mathcal{N},
(\bar{c}_{lj}, l \in \mathcal{L}, j \in \mathcal{N}))$. But
$\mathbf{m^{\ast}}$ may not be system optimal~(see
Example~\ref{ex-cont-opt2}).

\begin{remark}
1) $\bar{c}_{lj} = c_{lj} + t_{lj}$ can be interpreted as the
marginal cost due to additional association of class $l$ mobiles to
BS $j$. The term $c_{lj}$ is the power density incurred by these new
mobiles, and $t_{lj}$ is the increase in power consumption densities
of the mobiles already associated with BS $j$, integrated over all
such mobiles. Economists call them ``private cost'' and ``social
cost'', respectively. Selfish mobiles do not care for the social
cost, while the social optimality criterion accounts for this
marginal
externality~\cite{gametheory-comnet.roughgarden02selfish-routing}.

2) The cost functions for various classes have a certain structure
in the settings of interest to us. Mobile classes that consider a BS
pay tolls proportional to their required SINR densities. In
particular, tolls are uniform across all the mobile classes that
have equal SINR requirements. This is special to our setting;
usually one does not see uniform tolls in the case of multiclass
networks~(see
Dafermos~\cite{gametheory.dafermos73traffic-assignment-multi-class},
Smith~\cite{gametheory.smith79marginal-cost-taxation}).
\end{remark}

This toll mechanism can be implemented in a distributed fashion. All
the BSs broadcast the tolls~(normalized by SINR densities) along
with their aggregate congestions as before.\footnote{Normalized
tolls $\frac{t_{lj}}{\gamma_l}$ are uniform across all mobile
classes that consider a BS. A mobile can recover the exact toll from
the normalized value.} All mobiles need to know their scaled gains
$\frac{h_{lj}}{\sigma^2}$ to each BS $j \in \mathcal{N}$. A mobile
then makes a choice taking both power density and toll into account.

\subsection{Discrete Mobiles}
Pricing mechanisms for networks with discrete mobiles are relatively
difficult to design and analyze~(Fotakis \&
Spirakis~\cite{gametheory-comnet.fotakis-spirakis07cost-balancing-tolls}).
Again, we propose a toll mechanism that weakly enforces system
optimality in all cases and strongly enforces it in a special
setting. The mechanism is motivated by the toll mechanism for the
nonatomic case~(Theorem~\ref{opt-nonatom}).

Consider the network model of Section~\ref{sys-model} and an association profile $\mathbf{a'}$.
Let mobile $i$ evaluate BS $j$ for association.
Define $\mathbf{a} = (j, \mathbf{a'}_{-i})$.
Analogous to the nonatomic case,  define ``private'' and ``social'' costs as
\begin{eqnarray}
c_i(\mathbf{a}) &=& \frac{\sigma^2}{h_{ij}} \frac{\beta_i}{[1 - \sum_{k \in \mathcal{M}_j(\mathbf{a})}\beta_k]^+}, \nonumber \\
\mbox{and } t_i(\mathbf{a}) &=&  \sum_{l \in \mathcal{M}_j(\mathbf{a}) \setminus \{i\}}\frac{\sigma^2}{h_{lj}} \left( \frac{\beta_l}{[1 - \sum_{k \in \mathcal{M}_j(\mathbf{a})}\beta_k]^+} \right. \nonumber \\
&& \ \ \ \ \ \ \ \ - \left. \frac{\beta_l}{[1 - \sum_{k \in
\mathcal{M}_j(\mathbf{a})  \setminus \{i\}}\beta_k]^+}\right),
\label{eq-disc-toll}
\end{eqnarray}
respectively.\footnote{In~(\ref{eq-disc-toll}), when both terms
within parentheses are $\infty$, the expression is taken to be
$\infty$; we may think of driving $\beta$ to the true values from
below, and the first term always dominates the second. Same remark
holds for other such expressions also.} Clearly, $c_i(\mathbf{a})$
is the required power of mobile $i$ if it joins BS $j$, while
$t_i(\mathbf{a})$ is the aggregate increase in power consumption of
all other mobiles associated with BS $j$. We propose a toll
mechanism with tolls $t_i: \mathcal{N}^M  \rightarrow \mathbb{R}$
given by~(\ref{eq-disc-toll}). This yields a new game $(\mathcal{M},
\mathcal{N}, (\bar{c}_i, i \in \mathcal{M}))$ with cost functions
for an association profile $\mathbf{a}$ given by
\begin{eqnarray}
\hspace{-0.3in} \bar{c}_i(\mathbf{a}) &=&  c_i(\mathbf{a}) + t_i(\mathbf{a}) \nonumber \\
                      &=& \sum_{l \in \mathcal{M}_{a_i}(\mathbf{a})} \frac{\sigma^2}{h_{la_i}} \frac{\beta_l}{[1 - \sum_{k \in \mathcal{M}_{a_i}(\mathbf{a})}\beta_k]^+} - \nonumber \\
&& \sum_{l \in \mathcal{M}_{a_i}(\mathbf{a}) \setminus \{i\}}
\frac{\sigma^2}{h_{la_i}} \frac{\beta_l}{[1 - \sum_{k \in
\mathcal{M}_{a_i}(\mathbf{a}) \setminus \{i\}}\beta_k]^+}.
\end{eqnarray}
\begin{proposition}
The finite strategic form game $(\mathcal{M}, \mathcal{N}, (\bar{c}_i, i \in \mathcal{M}))$ is an ordinal potential game and thus admits FBRP.
\end{proposition}
\begin{IEEEproof}
For the game $(\mathcal{M}, \mathcal{N}, (\bar{c}_i, i \in \mathcal{M}))$,
the function $V:\mathcal{N}^{|\mathcal{M}|} \rightarrow \mathbb{R}$ given by
\[V(\mathbf{a}) = \sum_{i \in \mathcal{M}} \frac{\sigma^2}{h_{ia_i}} \frac{\beta_i}{[1 - \sum_{k \in \mathcal{M}_{a_i}(\mathbf{a})}\beta_k]^+}\]
is an ordinal potential function, as can be straightforwardly
checked. Thus $(\mathcal{M}, \mathcal{N}, (\bar{c}_i, i \in
\mathcal{M}))$ is an ordinal potential game. Since it is also a
finite game, the FBRP property holds.
\end{IEEEproof}

Note that the potential function $V(\mathbf{a})$ equals the system
performance measure $C(\mathbf{a})$ defined in
Section~\ref{sys-opt}. Hence an association profile $\mathbf{a}^o$
that optimizes system performance is also a (global) minimizer of
$V(\mathbf{a})$, and therefore a NE of the potential game with
tolls. \emph{So, we see that tolls $t_i(\mathbf{a})$ weakly enforce
a system optimal association profile.} In general, tolls do not
strongly enforce a system optimal association profile. For instance
reconsider Example~\ref{ex-sys-opt}. The association profile $(a_1 =
2, a_2 = 1)$ is inefficient, but an
 NE for the game $(\mathcal{M}, \mathcal{N}, (\bar{c}_i, i \in \mathcal{M}))$.

In the following we consider special cases, and investigate the effect of the proposed
tolls.

\subsubsection{Collocated Mobiles with Single Class Traffic}
Let us consider the special case when all the mobiles are collocated
and have identical minimum SINR requirements. In other words,
$h_{ij} = h_j$ and $\beta_i = \beta$  for all $i \in \mathcal{M}, j
\in \mathcal{N}$. The potential function for this special case can
be written as
\[V(\mathbf{a}) = \sum_{j \in \mathcal{N}} \frac{\sigma^2}{h_j} \frac{|\mathcal{M}_j(\mathbf{a})|\beta}{[1 - |\mathcal{M}_j(\mathbf{a})|\beta]^+}\]
Define $g_j = \frac{\sigma^2}{h_j}, f(m) = \frac{m\beta}{[1 -
m\beta]^+}$ and $m_j(\mathbf{a}) = |\mathcal{M}_j(\mathbf{a})|$ for
all $j \in \mathcal{N}$. Then $\mathbf{m}(\mathbf{a}) =
(m_j(\mathbf{a}), j \in \mathcal{N})$ denotes the congestion profile
under $\mathbf{a}$. Since mobiles are indistinguishable, any two
association profiles that lead to identical congestion profiles are
essentially indifferent from the point of view of analysis. Thus we
talk solely in terms of congestion profiles. Abusing notation~(the
argument of $V(\cdot)$ was earlier defined to be the association
profile $\mathbf{a}$), we write
\[V(\mathbf{m}) = \sum_{j \in \mathcal{N}} g_j f(m_j). \]
Since $(\mathcal{M}, \mathcal{N}, (\bar{c}_i, i \in \mathcal{M}))$
is a finite potential game, an association profile
$\mathbf{m}^{\ast}$ will be a NE if and only if
\begin{equation}
g_j f(m^{\ast}_j) + g_k f(m^{\ast}_k) \leq g_j f(m^{\ast}_j-1) + g_k
f(m^{\ast}_k+1) \label{price-NE-cond}
\end{equation}
for all $k \neq j, \ j,k \in  \mathcal{N}$. The following
proposition shows that tolls  $t_j(\mathbf{a})$ strongly enforce a
system optimal association profile in case of collocated mobiles
with single class traffic.

\begin{proposition}
All the NEs in the game $(\mathcal{M}, \mathcal{N}, (\bar{c}_i, i
\in \mathcal{M}))$, with $h_{ij} = h_j$ and $\beta_i = \beta$ for
all $i \in \mathcal{M}, j \in \mathcal{N}$, are system optimal. In
other words, the tolls strongly enforce system optimality.
\end{proposition}
\begin{IEEEproof}
Let  $\mathbf{m}^o$ be a system optimal congestion profile, and
$\mathbf{m}^{\ast}$ any other profile such that
$V(\mathbf{m}^{\ast}) > V(\mathbf{m}^o)$. Partition the set
$\mathcal{N}$ as $\mathcal{N} = \mathcal{N}_0 \cup \mathcal{N}_+
\cup  \mathcal{N}_-$ such that
\begin{eqnarray*}
j \in \mathcal{N}_0 &\Longleftrightarrow& m^{\ast}_j = m^o_j \\
j \in \mathcal{N}_+ &\Longleftrightarrow& m^{\ast}_j \geq m^o_j + 1 \\
j \in \mathcal{N}_- &\Longleftrightarrow& m^{\ast}_j \leq m^o_j - 1
\end{eqnarray*}

Start with the congestion profile $\mathbf{m}^{\ast}$, and  move
mobiles from BSs $\mathcal{N}_+$ to BSs $\mathcal{N}_-$ one mobile
at a time, so that we end up with the congestion profile
$\mathbf{m}^o$. In this process we get a succession of congestion
profiles, each of which satisfies
\begin{eqnarray*}
m_j = m^{\ast}_j  &\forall& j \in \mathcal{N}_0 \\
m_j \leq m^{\ast}_j  &\forall& j \in \mathcal{N}_+ \\
m_j \geq m^{\ast}_j  &\forall& j \in \mathcal{N}_-
\end{eqnarray*}

There must exist a pair of successive congestion profiles
$\mathbf{m}'$ and $\mathbf{m}''$ such that $V(\mathbf{m}') >
V(\mathbf{m}'')$, with $\mathbf{m}''$ possibly the ultimate
congestion profile $\mathbf{m}^o$. Let $\mathbf{m}''$ be obtained
from $\mathbf{m}'$ by the transfer of a mobile from BS $j \in
\mathcal{N}_+$ to a BS $k \in \mathcal{N}_-$. We then have
\[g_j f(m'_j) + g_k f(m'_k) > g_j f(m'_j-1) + g_k f(m'_k+1) \]
which is same as
\begin{equation} g_j(f(m'_j) - f(m'_j-1)) >
g_k(f(m'_k+1) - f(m'_k)). \label{eq-inter-cong}
\end{equation}
Recall that $f$ is a convex function and  $m'_j \leq m^{\ast}_j,
m'_k \geq m^{\ast}_k$. Using these in~(\ref{eq-inter-cong}), we get
\[g_j(f(m^{\ast}_j) - f(m^{\ast}_j-1)) > g_k(f(m^{\ast}_k+1) - f(m^{\ast}_k)),\]
i.e.,
\[g_j f(m^{\ast}_j) + g_k f(m^{\ast}_k) > g_j f(m^{\ast}_j-1) + g_k f(m^{\ast}_k+1)\]
which implies that $\mathbf{m}^{\ast}$ is not a
NE~(see~\eqref{price-NE-cond}). This completes the proof.
\end{IEEEproof}

\subsubsection{Collocated Mobiles and Symmetrically Placed BSs}
Now we consider another special case when all the mobiles are
collocated and all the BSs are symmetrically placed with respect to
the collocated mobiles. In this case $h_{ij} = h$ for all $i \in
\mathcal{M}, j \in \mathcal{N}$. We have the following result.

\begin{proposition}
With $h_{ij} = h$ for all $i \in \mathcal{M}, j \in \mathcal{N}$,
the NEs in the game $(\mathcal{M}, \mathcal{N}, (\bar{c}_i, i \in
\mathcal{M}))$ coincide with those in $(\mathcal{M}, \mathcal{N},
(c_i, i \in \mathcal{M}))$.
\end{proposition}
\begin{IEEEproof}
With $h_{ij} = h$ for all $i \in \mathcal{M}, j \in \mathcal{N}$,
\[
c_i(\mathbf{a}) = \frac{\sigma^2}{h} \frac{\beta_i}{[1 - \sum_{l \in
\mathcal{M}_{a_i}(\mathbf{a})}\beta_l]^+}.
\]
Thus an association profile $\mathbf{a}$ is a NE of the game
$(\mathcal{M}, \mathcal{N}, (c_i, i \in \mathcal{M}))$ if and only
if
\begin{equation}
\sum_{l \in \mathcal{M}_{a_i}(\mathbf{a})}\beta_l \leq \sum_{l
\in\mathcal{M}_j(\mathbf{a})} \beta_l + \beta_i \label{eqn:NE-char}
\end{equation}
for all $j \in \mathcal{N} \setminus \{a_i\}$ and $i \in
\mathcal{M}$. Next,
\begin{eqnarray*}
\bar{c}_i(\mathbf{a}) &=& \frac{\sigma^2}{h}\left(\frac{\sum_{l \in
\mathcal{M}_{a_i}(\mathbf{a})}\beta_l}{[1 - \sum_{l \in
\mathcal{M}_{a_i}(\mathbf{a})}\beta_l]^+}  \right. \\
&& \ \ \ \ \ \left. -  \frac{\sum_{l \in
\mathcal{M}_{a_i}(\mathbf{a})}
\beta_l - \beta_i}{[1 - \sum_{l \in \mathcal{M}_{a_i}(\mathbf{a}) \setminus \{i\}}\beta_l]^+} \right)\\
&=& \frac{\sigma^2}{h} \frac{\beta_i}{[1 - \sum_{l \in
\mathcal{M}_{a_i}(\mathbf{a})}\beta_l]^+ [1 - \sum_{l \in
\mathcal{M}_{a_i}(\mathbf{a}) \setminus \{i\}}\beta_l]^+}.
\end{eqnarray*}
It can be easily checked that NEs of the game $(\mathcal{M},
\mathcal{N}, (\bar{c}_i, i \in \mathcal{M}))$ are also characterized
by~\eqref{eqn:NE-char}. Hence we get the claim.
\end{IEEEproof}
Thus tolls may not strongly enforce a system optimal association profile
in this case~(see Example~\ref{ex-col-mobiles-BSs}).

\subsubsection{Collocated BSs with Single Class Traffic}
Even in this special case tolls $t_j(\mathbf{a})$ may fail to strongly enforce a system optimal association profile.
For an illustration reconsider Example~\ref{ex-col-BSs-single-class}. The association profile
 $(a_1 = a_3 = 1, a_2 = a_4  = a_5 = 2)$ is not system optimal, but an
 NE for the game $(\mathcal{M}, \mathcal{N}, (\bar{c}_i, i \in \mathcal{M}))$.

\begin{remark}
1) While tolls at a BS are equal for all the mobiles not associated
with it and having equal SINR requirements, they are mobile
dependent for all associated ones~(see~(\ref{eq-disc-toll})). This
is unlike in nonatomic case where we saw uniform tolls at a BS for
all the mobiles with equal SINR requirements.

2) The modified algorithm (the one accounting for tolls) can be
implemented in distributed fashion. All the BSs broadcast quantities
$t^o_j(\mathbf{a})$ given by
\[t^o_j(\mathbf{a}) = \sum_{l \in \mathcal{M}_j(\mathbf{a})} \frac{\sigma^2}{h_{lj}} \frac{\beta_l}{[1 - \sum_{k \in \mathcal{M}_j(\mathbf{a})}\beta_k]^+}\]
along with their aggregate congestions $\sum_{k \in
\mathcal{M}_j(\mathbf{a})}\beta_k$. All the mobiles need to know the
scaled gains $\frac{h_{ij}}{\sigma^2}$ of their own channels to all
the BSs $j \in \mathcal{N}$. Mobiles use these broadcast information
to calculate their powers and tolls, and choose a BS taking both
into account.
\end{remark}

\noindent
\paragraph*{Discussion}
The proposed pricing technique can be used to induce a system
optimal routing in atomic weighted network congestion games with
arbitrary nondecreasing edge latency
functions~\cite{gametheory-comnet.singh11pricing-atomic}.\footnote{Here,
the system cost is weighted sum of the latencies of all the users.}
In this setting, the joint BS association and power control problems
can be viewed as network congestion games over two-terminal
parallel-edge networks: the edges are identified with BSs, and
latencies are identified with minimum power requirements. It turns
out that the proposed tolls weakly enforce a system optimal routing
profile in general network congestion games. They strongly enforce a
system optimal routing profile if
\begin{enumerate}
\item the network is two-terminal series parallel,
\item the users are unweighted~(i.e, have identical weights), and
\item the latency functions are standard.\footnote{A latency function $c(\cdot)$ is called {\it standard}
if $m c(m)$ is
convex~\cite{gametheory-comnet.roughgarden02selfish-routing}, e.g.,
$c(m) = \frac{1}{1-m}$.}
\end{enumerate}

\section{Conclusion}
\label{future-work}
We studied the combined association and power control problem  in
multichannel multicell cellular networks. We studied the cases of discrete mobiles
and a continuum of mobiles. We proposed several distributed mechanisms motivated by the techniques of game theory.
We studied the inefficiency of the distributed algorithm in the case of a continuum of mobiles.
It is an open question whether such inefficiency can be quantified in the case of discrete mobiles.
To mitigate the inefficiency, we proposed toll mechanisms in both the settings.

Throughout we assumed only one BS per channel. It would be of interest to extend the work to multiple BSs per channel.


\appendices

\section{Nonatomic Power Control}
\label{nonatomic-pc} Assume $\mathcal{M}$ to be an infinite set of
mobiles, and a single BS.  The required SINRs of mobiles are given
by the function $\gamma : \mathcal{M} \rightarrow \mathbb{R}_{++}$.
Let $h: \mathcal{M} \rightarrow \mathbb{R}_{++}$ be the gains of
mobiles and let $p : \mathcal{M} \rightarrow \mathbb{R}_{++}$ be the
power allocation. The functions $p(x)$ and $\gamma(x)$ are
interpreted as power density and target SINR density, respectively,
per unit mass. The feasibility condition for $p$ can be written as
\[
\frac{p(x)h(x)}{\int_{\mathcal{M}}p(y)h(y)dy + \sigma^2} \geq \gamma(x), \ \forall x \in \mathcal{M},
\]
or equivalently,
\[
p(x)h(x) \geq \gamma(x)\int_{\mathcal{M}}p(y)h(y)dy + \gamma(x)\sigma^2, \ \forall x \in \mathcal{M}.
\]
Integrating the inequalities over the set of all mobiles, we get the following necessary condition for
feasibility.
\begin{eqnarray}
\label{nec-cond}
\int_{\mathcal{M}}p(x)h(x)dx &\geq& \int_{\mathcal{M}}\gamma(x)dx \int_{\mathcal{M}}p(y)h(y)dy \nonumber \\
&& + \ \sigma^2 \int_{\mathcal{M}}\gamma(x)dx \\
                             &>& \int_{\mathcal{M}}\gamma(x)dx \int_{\mathcal{M}}p(y)h(y)dy \nonumber
\end{eqnarray}
where the strict inequality arises because $\sigma^2 \int_{\mathcal{M}}\gamma(x)dx > 0$.
Canceling $\int_{\mathcal{M}}p(y)h(y)dy$ from both sides, we get the necessary condition
\begin{equation}
\label{feas-cond}
\int_{\mathcal{M}}\gamma(x)dx < 1.
\end{equation}
Assuming~\eqref{feas-cond} holds,
\begin{equation}
\label{power-alloc}
p(x) = \frac{\gamma(x)\sigma^2}{h(x)(1 - \int_{\mathcal{M}}\gamma(x)dx)}
\end{equation}
is a feasible power allocation, and so~\eqref{feas-cond} is
necessary and sufficient for feasibility of the power control
problem.

The power allocation given by~\eqref{power-alloc} is Pareto efficient.
Let us see why.
Suppose  $q : \mathcal{M} \rightarrow \mathbb{R}_{++}$ is another feasible power allocation such that
$q(x) \leq p(x)$ and strict inequality holds for a set of mobiles having a positive measure, and so
\begin{eqnarray*}
\int_{\mathcal{M}}q(x)h(x)dx &<& \int_{\mathcal{M}}p(x)h(x)dx,\\
                             &=&  \frac{\sigma^2 \int_{\mathcal{M}}\gamma(x)dx}{1 - \int_{\mathcal{M}}\gamma(x)dx}
\end{eqnarray*}
where the equality uses~\eqref{power-alloc}. Thus,
$q$ violates the necessary condition
\[
\int_{\mathcal{M}}q(x)h(x)dx \geq \frac{\sigma^2 \int_{\mathcal{M}}\gamma(x)dx}{1 - \int_{\mathcal{M}}\gamma(x)dx},
\]
a rearrangement of (\ref{nec-cond}), and hence cannot be feasible.

In fact $p$ can be shown to be the unique Pareto efficient and hence the system optimal
 power allocation.\footnote{The system optimality criterion is
to minimize the sum of power consumptions over all the mobiles.}
Indeed if $q$ is another feasible power vector which is also Pareto
efficient,  the pointwise minimizer of $p$ and $q$ is also feasible
and Pareto dominates $p$, thus contradicting the fact that $p$ is
Pareto efficient.

\section{Price of Anarchy: Continuum of Mobiles}
Recall that a NE is not necessarily a system optimal congestion
profile~(see Example~\ref{ex-cont-opt}). {\it Coordination
ratio}~\cite{gametheory.koutsoupias-papadimitriou} or {\it Price of
Anarchy}~\cite{gametheory-wireless.papadimitrio-algorithms-games-internet}
characterizes the inefficiency caused by the selfish behavior of
players; it is the ratio of the cost of the worst NE and the optimal
cost. We observed in Proposition~\ref{cost-unique} that, in the
nonatomic case, mobiles incur the same cost at all the NEs. We can
then define price of anarchy as follows.

\begin{definition}
Let $\mathbf{m}$ be a NE, and $\mathbf{m}^o$ be a system optimal
congestion profile. Then the {\it price of anarchy} is
\[
\PoA = \frac{C(\mathbf{m})}{C(\mathbf{m}^o)}.
\]
\end{definition}

We restrict our analysis to a single class population. We assume
that all the mobiles have identical minimum required SINR density
$\gamma$ and identical power gain $h_j$ to BS $j$, $j \in
\mathcal{N}$.

\subsection{Two BSs}
First we consider a case with $2$ BSs as in
Example~\ref{ex-cont-opt}. Let $h_1 > h_2$.\footnote{If $h_1 = h_2$
equal fraction of population join each of the BSs under the NE and
the system optimal association, and the price of anarchy is $1$.}
Also, let $(\alpha^{\ast} M,(1-\alpha^{\ast})M)$ and $(\alpha^o
M,(1-\alpha^o)M)$ be the congestion profiles under a NE and a system
optimal association, respectively. Recall from
Example~\ref{ex-cont-opt} that
\begin{enumerate}
\item if $\gamma M \leq 1 - \sqrt{\frac{h_2}{h_1}}$, then $\alpha^{\ast} = \alpha^o =1$
\item if $1 - \sqrt{\frac{h_2}{h_1}} < \gamma M \leq 1 - \frac{h_2}{h_1}$, then $\alpha^{\ast} = 1$,
and from~\eqref{eq-cont-opt}
\[\alpha^o = \frac{\sqrt{h_1}-\sqrt{h_2}+\gamma M \sqrt{h_2}}{(\sqrt{h_1} + \sqrt{h_2}) \gamma M}\]
\item if $\gamma M > 1 - \frac{h_2}{h_1}$, then from~\eqref{eq-cont-nash}
\[\alpha^{\ast} = \frac{h_1-h_2+\gamma M h_2}{(h_1 + h_2) \gamma M},\]
and $\alpha^o$ is as above.
\end{enumerate}
$C(\mathbf{m})$ and $C(\mathbf{m}^o)$ are obtained via substituting
$\alpha = \alpha^{\ast}$ and $\alpha= \alpha^o$, respectively, in
the objective function~\eqref{exm-obj-function}. Straightforward
calculations give that
\remove{
\begin{equation*}
\PoA(M) = \left\{ \begin{array}{llll}
                1 \mbox{ if } M \leq \frac{1}{\gamma}\left(1 - \sqrt{\frac{h_2}{h_1}}\right), \\
        \frac{h_2(2- \gamma M)\gamma M}{(1 - \gamma M)(2 \sqrt{h_1 h_2} - (h_1 + h_2)(1 - \gamma M))} \\
        \ \ \ \ \ \ \ \mbox{ if } \frac{1}{\gamma}\left(1 - \sqrt{\frac{h_2}{h_1}}\right) \leq M \leq \frac{1}{\gamma}\left(1 - \frac{h_2}{h_1}\right), \\
                \frac{\gamma M (h_1 + h_2)}{2\sqrt{h_1 h_2} - (h_1 + h_2)(1 - \gamma M)} \mbox{ if } M \geq \frac{1}{\gamma}\left(1 - \frac{h_2}{h_1}\right).
                     \end{array} \right.
\end{equation*}
Defining $\lambda := \frac{h_2}{h_1} < 1$, we get
}
\begin{equation*}
\PoA(M) = \left\{ \begin{array}{llll}
                1 \ \ \ \ \ \ \ \ \ \ \ \ \ \ \ \ \ \ \ \ \  \ \ \ \ \mbox{ if } M \leq \frac{1 - \sqrt{\lambda}}{\gamma}, \\
        \frac{\lambda(2- \gamma M)\gamma M}{(1 - \gamma M)(2 \sqrt{\lambda} - (1 - \gamma M)(1 + \lambda))} \\
        \ \ \ \ \ \ \ \ \ \ \ \ \ \ \ \ \ \ \  \ \ \ \ \ \ \ \mbox{ if } \frac{1 - \sqrt{\lambda}}{\gamma} \leq M \leq \frac{1 - \lambda}{\gamma}, \\
                \frac{\gamma M (1 + \lambda)}{2\sqrt{\lambda} - (1 - \gamma M)(1 + \lambda)} \ \ \ \ \ \mbox{ if } M \geq \frac{1 - \lambda}{\gamma}
                     \end{array} \right.
\end{equation*}
where $\lambda := \frac{h_2}{h_1} < 1$. Further calculations also
yield that $\PoA(M)$ is continuous at $M = \frac{(1 -
\lambda)}{\gamma}$, and
\begin{equation*}
\frac{{\rm d}\PoA(M)}{{\rm d}M} \left\{ \begin{array}{ll}
                    \geq 0 \mbox{ if } M < \frac{1 - \lambda}{\gamma},\\
                    \leq 0 \mbox{ if } M > \frac{1 - \lambda}{\gamma}
                               \end{array} \right.
\end{equation*}
Thus, the price of anarchy is maximized when $M = \frac{1 -
\lambda}{\gamma}$. Moreover, the maximum price of anarchy is
\[
\frac{1 - \lambda^2}{2\sqrt{\lambda} - \lambda(1 + \lambda)}.
\]
Viewing this now as a function of $\lambda \in (0,1]$, we see that
the maximum price of anarchy decreases with $\lambda$. We also
observe that $\PoA \rightarrow \infty$ as $\lambda \rightarrow 0$,
i.e., arbitrarily high PoAs can be realized in $2$ BS networks.

\subsection{$N$ BSs}
\begin{figure}[h]{
\centering
\includegraphics[height=2.5in]{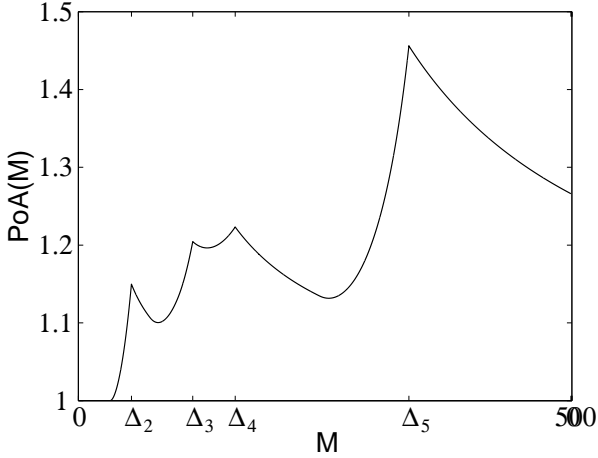}
\caption{PoA vs $M$: the price of anarchy is maximized at one of the masses $\{\Delta_2,\Delta_3,\Delta_4,\Delta_5\}$.}
 \label{fig:PoA}}
\end{figure}

Again, without any loss of generality, we assume that $h_1 \geq h_2
\geq \cdots \geq h_N$. We also assume that the population's mass is
$\Delta_j$ when it spills over BS $j$ under NE. Clearly, $\Delta_2
\leq \Delta_3 \leq \cdots \leq \Delta_N$. In the case of $2$ BSs we
proved that price of anarchy is maximized when the population spills
over BS $2$ under NE. In the case of $N > 2$ BSs also, simulations
suggest that the price of anarchy is maximized  at one of the spill
over points $\{\Delta_j, j = 2,\dots,N\}$. We have however not been
able to prove this observation. We illustrate this observation in
Figure~\ref{fig:PoA} which shows the price of anarchy as a function
of mass of the population in a network with $5$ BSs. For this
example $\gamma = 0.01$, and the power gains are chosen
independently and uniformly from $[0,1]$.

In the following, we give expression for the price of anarchy when
the total population has mass $\Delta_j, j \geq 2$. Let $m_k, k <
j$, be the mass of the population associated with BS $k$ under NE.
From the definition of NE~(see Definition~\ref{cont-player-NE}),
\[
\frac{\gamma \sigma^2}{h_k(1 - \gamma m_k)} = \frac{\gamma \sigma^2}{h_j} \mbox{ for all } k < j.
\]
Thus,
\[
\Delta_j  = \sum_{k < j}m_k = \frac{j-1}{\gamma} - \frac{h_j e_{j-1}}{\gamma}
\]
where $e_j := \sum_{k \leq j}\frac{1}{h_k}$ for $j \geq 1$.
The cost at NE
\[
C(\mathbf{m}) = \sum_{k < j}m_k \frac{\gamma \sigma^2}{h_j} = \sigma^2 \left(\frac{j-1}{h_j} - e_{j-1}\right).
\]

Next, consider the system optimal congestion profile
$\mathbf{m^{\ast}}$. It can be easily checked that $m^{\ast}_k$ is
positive only if $k < j'$ for a $j < j' \leq N$. Moreover, it must
be the case that
\begin{eqnarray*}
\sqrt{h_k}(1 - \gamma m^{\ast}_k) &=& K \mbox{ for all } k < j',\\
\mbox{and } \sqrt{h_{j'}} &\leq&  K,
\end{eqnarray*}
for some constant $K >0$. Recall that $\sum_{k < j'}m^{\ast}_k = \Delta_j$.
Straightforward calculations yield that
\begin{equation*}
K = \frac{j'-j + h_j e_{j-1}}{e^{\ast}_{l-1}}
\end{equation*}
where $e^{\ast}_j := \sum_{k \leq j}\frac{1}{\sqrt{h_k}}$ for $j \geq 1$.
Moreover, $j'$ is the least index satisfying
\[
\sqrt{h_{j'}} \leq \frac{j'-j + h_j e_{j-1}}{e^{\ast}_{l-1}}.
\]
The optimal cost, after substituting the values of $m^{\ast}_k$, turns out to be
\begin{eqnarray*}
C(\mathbf{m^{\ast}}) &=& \sum_{k < j'} m^{\ast}_k \frac{\gamma \sigma^2}{h_k(1 - m^{\ast}_k \gamma)} \\
                     &=& \sigma^2\left(\frac{e^{{\ast}^2}_{j'-1}}{j'- j + h_j e_{j-1}} - e_{j'-1}\right).
\end{eqnarray*}
The price of anarchy $\PoA(\Delta_j)$ is the ratio $\frac{C(\mathbf{m})}{C(\mathbf{m^{\ast}})}$.

Finally, we show that the price of anarchy decreases with mass for $M \geq \Delta_N$.
It can be easily checked that, for $M \geq \Delta_N$,
\begin{eqnarray*}
\PoA(M) &=& \frac{e_N M \gamma}{e_N M \gamma - (e_N N - e^{{\ast}^2}_N)} \\
       &=& 1 + \frac{e_N N - e^{{\ast}^2}_N}{e_N M \gamma - (e_N N - e^{{\ast}^2}_N)}
\end{eqnarray*}
from which the claim follows. Thus, to obtain a bound on the price of anarchy, we only focus on $M \leq \Delta_N$.
For $M \leq \Delta_N$, the load on BS~$j$
\[m_j \leq \frac{1}{\gamma}\left(1- \frac{h_N}{h_j}\right)\]
under NE. We use this observation in the next section.

\subsection{A Bound on the Price of Anarchy}
Now, we derive a sharp bound on the price of anarchy for single class networks with arbitrary number of
BSs, and gains $h_j \in [h_{\min},h_{\max}]$ for all the BSs. We follow
Roughgarden~\cite[Chapter~$3$]{gametheory-comnet.roughgarden02selfish-routing}.

In the BS association game, a generic cost function is of the form
\[
c_h(m) := \frac{\sigma^2}{h} \frac{\gamma}{1 - \gamma M},
\]
and
\[\mathcal{C} := \{c_h(\cdot): h \in [h_{\min},h_{\max}]\}
\]
is the class of all feasible cost functions. Observe that the
functions $c_h(\cdot)$ and the class $\mathcal{C}$ both are
standard.\footnote{A cost function $c(\cdot)$ is called {\it
standard} if $m c(m)$ is convex. A class $\mathcal{C}$ is {\it
standard} if it contains a nonzero function and if each $c(\cdot)
\in \mathcal{C}$ is
standard~\cite{gametheory-comnet.roughgarden02selfish-routing}.} We
define
\[\bar{c}_h(m) := \frac{{\rm d} (mc_h(m)}{{\rm d} m}.\]
We also assume that the load on a BS with gain $h_j$ does not exceed
\[\theta_h := \frac{1}{\gamma}\left(1- \frac{h_{\min}}{h}\right)\]
under NE. Thus, we redefine {\it anarchy value} for a cost function $c_h(\cdot)$ as\footnote{The original
definition~\cite[Definition~3.3.2]{gametheory-comnet.roughgarden02selfish-routing} considers supremum
over $m \in (0,\infty)$.}
\[
\alpha(c_h) := \sup_{m \leq \theta_h}[\lambda \mu + (1 - \lambda)]^{-1}
\]
where $\lambda \in (0,1)$ satisfies $\bar{c}_h(\lambda m) = c_h(m)$ and $\mu := \frac{c_h(\lambda m)}{c_h(m)} \leq 1$.
Both $\lambda$ and $\mu$ are functions of $m$; we do not show this dependence explicitly.
Straightforward calculations yield that
\begin{eqnarray*}
\lambda &=&  \frac{1- \sqrt{1 - m \gamma}}{m \gamma}, \\
\mu &=& \sqrt{1 - m \gamma}, \\
\mbox{and } \alpha(c_h) &=& \sup_{m \leq \theta_h}\frac{1}{2}\left[1 - \frac{1}{1 + \sqrt{1 - m \gamma}}\right]^{-1} \\
                        &=& \frac{1}{2}\left(1 + \sqrt{\frac{h}{h_{\min}}}\right).
\end{eqnarray*}
The {\it anarchy value} for class  $\mathcal{C}$ is~(see \cite[Definition~3.3.3]{gametheory-comnet.roughgarden02selfish-routing})
\[
\alpha(\mathcal{C}) = \sup_{c_h \in \mathcal{C}} \alpha(c_h) = \frac{1}{2}\left(1 + \sqrt{\frac{h_{\max}}{h_{\min}}}\right)
\]
It can be easily checked that~\cite[Theorem~3.3.8]{gametheory-comnet.roughgarden02selfish-routing}
remains valid with our new definition of anarchy value. Thus, price of anarchy is bounded by
$\alpha(\mathcal{C})$. For any $0 < \epsilon \leq \alpha(\mathcal{C}) - 1$, a price of anarchy $\geq \alpha(\mathcal{C}) - \epsilon$
is realized in a network in which~$(i)$ there is one BS with gain $h_{\max}$,~$(ii)$ there are
several BSs with gain $h_{\min}$~(minimum number depending on $\epsilon$), and~$(iii)$ the population has
mass $\theta_{h_{\max}}$~(see the proof of~\cite[Lemma~3.4.3]{gametheory-comnet.roughgarden02selfish-routing} for details).

\bibliographystyle{IEEEtran}
\bibliography{IEEEabrv,ctrl-theory,game-theory,comm-net}

\end{document}